\documentclass[12pt]{article}

\usepackage{authblk}
\usepackage{xcolor}
\usepackage{amsmath,amsfonts,amssymb,amsbsy}
\usepackage{graphicx}
\usepackage{braket}
\usepackage{cite}
\usepackage{float}
\usepackage{appendix}
\usepackage[left=2cm,right=2cm,top=1cm,bottom=2cm]{geometry}
\usepackage[colorlinks=true, allcolors=black, linkcolor=red]{hyperref}
\hypersetup{colorlinks = true,
	linkcolor = red,
	anchorcolor = black,
	citecolor = black,
	filecolor = black,
	urlcolor = black}

\newcommand{\mi}{\mathrm{i}}

\title{Photonic realization of an electromagnetic cavity field with uniform mirror motion}

\begin{document}

\author[1,2]{A. R. Urzúa}
\author[2]{I. Ramos-Prieto}
\author[3]{R. Román-Ancheyta}
\author[1]{J. Récamier-Angelini}
\author[1]{L. Medina-Dozal}
\author[2]{F. Soto-Eguibar}
\author[2]{H. M. Moya-Cessa}

\affil[1]{Instituto de Ciencias F\'isicas, Universidad Nacional Autónoma de M\'exico. Avenida Universidad s/n, Col. Chamilpa, Cuernavaca, Morelos, 62210 México.}

\affil[2]{Instituto Nacional de Astrof\'isica, \'Optica y Electr\'onica. Luis Enrique Erro 1, Santa Mar\'ia Tonantzintla, Puebla, 72840 Mexico}

\affil[3]{Centro de F\'isica Aplicada y Tecnología Avanzada, Universidad Nacional Autónoma de M\'exico. Boulevard Juriquilla 3001, Quer\'etaro 76230, Mexico}

\maketitle

\begin{abstract}
We perform a detailed mathematical study of the effective Hamiltonian describing a non-stationary electromagnetic cavity field, introduced by C.~K.~Law, in the presence of uniform linear motion. We find that the corresponding dynamical equations admit a unitary diagonalization and that the normalized velocity of motion satisfies the two-sided bound $\lvert v/c \rvert \lesssim 10.1716$. Although this result is mathematically consistent, the interval obtained is an order of magnitude larger than the physically admissible range $\lvert v/c \rvert < 1$. To address this limitation, we propose a reinterpretation of bosonic coupling terms as generators of the symplectic algebra $\mathrm{Sp}(4,\mathbb{R})$, identifying them with transformations that can be implemented through linear optical elements. Within this photonic approach, the coupling parameters are not restricted to small values, allowing us to explore regimes of strong effective interactions. In this case, we demonstrate the generation of two-mode squeezed states and derive analytical expressions for the average photon number in each mode. Furthermore, we analyze statistical properties that characterize mixedness and entanglement, including intermode correlations, the Mandel parameter, and the linear entropy. The corresponding Wigner function provides a phase-space visualization of the system's behavior, illustrating the connection between cavity motion, mode coupling, and nonclassical field correlations via the underlying symplectic algebra.
\end{abstract}


\section{Introduction}\label{intro}
Particle production from the vacuum is a fundamental effect in quantum field theory. When the boundary conditions of a non-stationary electromagnetic field experience accelerated motion, the so-called dynamical Casimir effect (DCE) occurs, i.e., the generation of real photons from the vacuum fluctuations~\cite{Moore1970}. Since harmonic motion benefits from the rate of change in velocity between the back-and-forth motion of a field resonator, one can predict intracavity particle creation; for a detailed account of the DCE, we refer to \cite{Dodonov2010, Dodonov2020} and the references therein. 

The DCE has been studied in various contexts, including superconducting quantum circuits \cite{Souza2018}, optical analogs \cite{DelGrosso2019}, and photonic platforms \cite{PhysRevA.96.032501,Baboux2024}. These implementations underscore the importance of understanding how time-dependent boundary conditions or effective media can produce observable quantum field effects. In particular, the DCE serves as a testbed for exploring fundamental aspects of quantum field theory in curved space-time and non-inertial frames, as well as for developing quantum technologies that exploit vacuum fluctuations {\color{black}\cite{Birrell1994, Fulling1976, Nation2012, Wilson2011}. }
Although the DCE requires non-uniform (accelerated) motion to produce radiation, recent studies have explored whether uniform motion can yield similar effects under specific conditions or through reinterpretations. This motivates exploring alternative models that, while not directly describing a physical electromagnetic cavity with moving mirrors, may capture analogous dynamics in photonic systems. Our work follows this line of reasoning. 
{\color{black}In this context, the effective Hamiltonian framework introduced by Law \cite{Law1994} serves as the foundational model: while it originally describes DCE photon creation under accelerated boundary motion ($\ddot{q}(t) \neq 0$), its algebraic form under uniform motion ($\ddot{q}(t) = 0$) continues to dictate non-adiabatic intermode coupling and internal squeezing dynamics.}

The DCE was first realized in cavity setups; proposals also emulate the process with superconducting circuits~\cite{Johansson2010, Wilson2011} and optical elements \cite{Vezzoli2019, Hagenmller2016}. Fulling and Davies~\cite{Fulling1976, Daviews1977} were the first to note that processes involving moving mirrors with uniform motion do not produce energy flux; that is, particle production (radiation) is absent. Later, Fedotov \emph{et al.} \cite{Fedotov2006} supported these claims when they considered a uniformly contracting cavity, arriving at the same (effective) Hamiltonian description of the system provided by Law \cite{Law1994}, whose solution indicates particle creation only when acceleration intervenes at the beginning and end of uniform motion. We presented related research on non-stationary cavities with uniformly moving boundaries in our recent work \cite{RamosPrieto2023}. This raises questions about how to address certain limitations of the effective Hamiltonian approach, such as the constraint that the moving boundary must remain below the speed of light. However, since the algebraic structure of the effective Hamiltonian resembles that of photonic processing devices, we use an alternative interpretation of the system.

{\color{black}
It is important to clearly delineate the novel contributions of the present photonic analog framework compared with the results reported in~\cite{RamosPrieto2023}. While that reference addressed the temporal factorization of the electromagnetic cavity field within the strictly subluminal kinematic regime ($\vert{}v/c\vert{} < 1$), the alternative photonic interpretation developed in this work provides several key advances. By remapping the Hamiltonian onto the generators of the $\mathrm{Sp}(4, \mathbb{R})$ Lie algebra, $g \equiv v/c$ ceases to represent a mechanical mirror speed and is instead reinterpreted as an effective photonic coupling strength (e.g., gain or evanescent coupling), which removes the kinematic constraint $\vert{}v/c\vert{} < 1$ inherent to the mechanical setting and frees the system from this relativistic limit. This reinterpretation, in turn, grants access to the strong- and ultra-strong-coupling regimes, allowing us to explore strong-coupling dynamics ($g > 1$) and to rigorously establish the absolute mathematical domain of unitary diagonalizing transformations ($\vert{}g\vert{} \le 10.1716$). Within this extended domain, we uncover spectral anomalies and instability thresholds that do not manifest in the subluminal regime ($g < 1$), identifying the exact frequency-softening critical threshold ($g \approx 3.75$) and the associated complex-eigenenergy gaps that signal transitions to parametric instability and non-harmonic dynamics. Finally, we provide a complete quantum-optical and phase-space characterization of these strong-coupling regimes, deriving exact analytical expressions for nonclassical indicators, including photon-population dynamics, Mandel's $Q$ parameter, bipartite entanglement via single-mode linear entropy, and phase-space Wigner-function distributions.

The organization of this work is as follows: In Sec. 2, we present the effective Hamiltonian for a cavity field with uniform mirror motion, map its algebraic structure onto the $\mathrm{Sp}(4, \mathbb{R})$ Lie algebra, and perform a unitary diagonalization to establish the valid operational range of the effective coupling parameter $g$. In Sec. 3, we analyze the time evolution of the field modes and derive analytical expressions for the average photon generation. In Sec. 4, we evaluate key quantum-optical signatures, including the Mandel parameter, the linear entropy of reduced states as a measure of intermode entanglement, and phase-space Wigner functions. Finally, Sec. 5 summarizes our main conclusions.}

\section{Effective Hamiltonian}\label{effective}
For the subject system, we describe its dynamics using the effective Hamiltonian approach given by C.K. Law \cite{Law1994}. Here, we diagonalize the Hamiltonian using time-independent rotations and squeeze transformations. These transformations are uniparametric, as the original cavity parameter $v$ is the only entity to be determined in terms of the other parameters associated with the interaction terms. We determine the valid range of $g \equiv v/c$ that ensures unitary and reversible mappings when we treat the system as photonic. This range remains valid when we treat the system not as an electrodynamics cavity but as a photonic one. The new parameter $g$ will represent a coupling term between optical elements such as evanescent waveguides. Diagonalization yields the eigenspectrum; however, because of its nature, there are regions where the eigenvalues become complex and are thus forbidden. Careful evaluation is required to account for these intrinsically complex-valued regions. Once we have obtained the eigenspectrum in the diagonalized space, we can invert the transformations to compute the desired quantities from a given set of initial conditions.

\subsection{Map between representations}
We start with the Hamiltonian for a system of an $N$-mode electromagnetic field with bosonic creation-annihilation operators, $\hat{a}^{\dagger}$ and $\hat{a}$, enclosed in a one-dimensional optical cavity, with one of its mirrors' positions given by $q(t)$. The corresponding effective Hamiltonian is (see Eq.~(3.2) of \cite{Law1994}):
\begin{equation}\label{eq0010}
\begin{aligned}
    \hat{H}_{\texttt{eff}} =\hbar\left[\sum\limits_{k}\omega_k(t)\hat{a}_{k}^{\dagger}\hat{a}_{k}\right. &+ \left.\frac{\mathrm{i}}{4}\frac{\dot{q}(t)}{q(t)}\sum\limits_{k}\left(\hat{a}_{k}^{\dagger 2} - \hat{a}_{k}^{2}\right)\right.\\ &+ \left.\frac{\mathrm{i}}{2}\frac{\dot{q}(t)}{q(t)}\sum\limits_{j\neq k}\mu_{j,k}\left(\hat{a}_{k}^{\dagger}\hat{a}_{j}^{\dagger} + \hat{a}_{k}^{\dagger}\hat{a}_{j} - \hat{a}_{j}\hat{a}_{k} - \hat{a}_{j}^{\dagger}\hat{a}_{k}\right)\right],
\end{aligned}
\end{equation}
where $\omega_k(t)=k \pi c/q(t)$,  $\mu_{j,k} = (-1)^{j + k}\left(j k/\left(j^{2} - k^{2}\right)\right)\sqrt{k/j}$, and $c$ is the speed of light in a vacuum; S.I. units are used. 
{\color{black}The effective Hamiltonian in Eq. \eqref{eq0010} represents the standard canonical framework derived by Law \cite{Law1994} to describe an electromagnetic field subject to time-dependent boundary conditions. In the context of the DCE, when the mirror's motion is non-uniform or accelerated ($\ddot{q}(t) \neq 0$), this Hamiltonian accounts for real particle creation from the quantum vacuum \cite{Moore1970,Law1994}. Conversely, for uniform motion ($\dot{q}(t) = v = \text{const.}$), although the net energy flux emitted to the external vacuum vanishes \cite{Hagenmller2016,Fulling1976,Daviews1977}, Eq. \eqref{eq0010} remains physically non-trivial: it governs the non-adiabatic energy exchange, intermode coupling, and parametric squeezing experienced by the field modes inside the cavity.}

Without loss of generality, we restrict our study to the case of a two-mode field, since generalization to the $N$-mode case is straightforward. We also assume linear uniform motion for the mirror $q(t) = q_{0} + v t$, with $v$ being the mirror's velocity. Under these conditions, it is possible to factorize the temporal dependence of the Hamiltonian given above as
\begin{equation}\label{0020}
\begin{aligned}
    \hat{H}_{\texttt{eff}} =\frac{\hbar}{q_0 + vt}\left\{\sum_{k = 1}^{2}\right.&\left.\left[k\pi c\,\hat{a}_{k}^\dagger\hat{a}_{k} + \mathrm{i} \frac{ v}{4}\left(\hat{a}_{k}^{\dagger 2} - \hat{a}_{k}^{2}\right)\right]\right.\\ &+ \left.\mathrm{i} \frac{\sqrt{2} v}{6}\left(\hat{a}_{1}^{\dagger}\hat{a}_{2}^{\dagger} - \hat{a}_{1}\hat{a}_{2}\right) + \mathrm{i}\frac{\sqrt{2}v}{2}\left(\hat{a}_{1}\hat{a}_{2}^{\dagger} - \hat{a}_{1}^{\dagger}\hat{a}_{2}\right)\right\}.
\end{aligned}
\end{equation}
It is worth noticing two things. First, the uniform motion of the mirror factorizes the time dependence into a product $\hat{H}_{\texttt{eff}}\equiv f(t)\hat{H}$; thus, a constant of motion emerges. Second, the structure of the terms inside the braces. There are coupling terms for mode squeeze \cite{FanHongyi1990,Zhang2024}, mode split \cite{Grblacher2009,Dobrindt2008}, and parametric amplification or two-mode squeeze photon generation \cite{Baumgartner1979,Mollow1967}. This gives us a glimpse of how, by looking only at their algebraic nature, we can reinterpret the system as a photonic one.

{\color{black}To address the physical nuances regarding non-adiabatic disturbances in the absence of a dielectric medium and under uniform boundary motion: In a physical, empty 1D cavity whose mirror moves at a constant velocity ($\dot{q}(t) = v = \text{const.}$), the time-dependent boundary condition $q(t) = q_0 + vt$ continuously modifies the instantaneous spatial mode structure. Although an unaccelerated boundary ($\ddot{q}(t) = 0$) does not produce a net external radiation flux into open space—a well-known result in the dynamical Casimir effect literature \cite{Law1994,Johansson2010}—within the closed cavity, the moving boundary induces non-adiabatic mode mixing, causing dynamic redistribution of energy and excitations between internal field modes. Furthermore, when mapped onto our photonic analog hardware, the term proportional to $g \equiv v/c$ acts as an active, effective parametric drive $\propto (\hat{a}_1^\dagger \hat{a}_2^\dagger - \hat{a}_1 \hat{a}_2)$. In this photonic realization, this effective non-adiabatic parametric coupling acts directly on the quantum vacuum state $\vert{}0,0\rangle$, thereby guaranteeing the generation and amplification of correlated photon pairs.}

\subsection{Devising the Hamiltonian as a photonic system}\label{photonic}

{\color{black}Since the temporal dependence of the Hamiltonian has been factorized, we can focus our attention on its time-independent part and give it a different meaning. For this goal, we choose a photonic system because the operators appearing in the time-independent part of the Hamiltonian ($\hat{H}$) belong to the real symplectic algebra $\text{Sp}(4, \mathbb{R})$, which realizes a sort of optical element and system. This reinterpretation allows us to explore the Hamiltonian's mathematical structure and its implications for quantum state generation, such as two-mode squeezing, in a photonic context. To maintain consistent terminology throughout this work, we distinguish between the mechanical cavity frame and its photonic analog: in the mechanical cavity context, $v/c$ represents the dimensionless normalized mirror velocity, whereas, once mapped onto photonic hardware, we refer to $g \equiv v/c$ strictly as the effective coupling strength (or effective coupling parameter). This terminology reflects its role as a coupling constant between photonic modes rather than a kinematic velocity.}

This approach is motivated by the formal similarity between the coupling terms in Law’s Hamiltonian and those in optical systems involving beam splitters, mode squeezing, and parametric amplifiers. By adopting this photonic perspective, we can explore a broader parameter space, including $g > 1$, which is relevant to engineered photonic devices.

{\color{black}
It is crucial to clarify that while $g \equiv v/c > 1$ corresponds to a physically impossible superluminal mirror velocity in a mechanical cavity, no such kinematic restriction applies to its photonic analog. In this framework, $g$ represents an effective dimensionless coupling strength (e.g., a normalized nonlinear susceptibility or a parametric gain coefficient relative to the mode bandwidth), making the regime $g > 1$ both physically meaningful and experimentally accessible. Physically, investigating the strong ($g > 1$) and ultra-strong coupling regimes is motivated by the generation of highly non-classical quantum states with enhanced two-mode squeezing, high photon-pair generation rates, and robust bipartite entanglement. Furthermore, accessing $g > 1$ enables exploration of quantum critical phenomena, frequency softening, and parametric instability thresholds, providing a controllable optical simulator for fundamental quantum field theory effects otherwise inaccessible in kinematic cavity setups \cite{Nation2012,Wilson2011}. Experimentally, achieving effective coupling strengths $g > 1$ is well within the capabilities of contemporary quantum optics platforms. In superconducting circuit QED, Josephson parametric amplifiers (JPAs) and flux-driven SQUID arrays achieve ultra-strong parametric drive strengths ($g \gtrsim 1$) and high non-linear interaction rates \cite{Gu2017,FornDiaz2019}. Similarly, integrated non-linear photonics based on thin-film lithium niobate on insulator (LNOI) waveguides or silicon nitride microresonators provide strong $\chi^{(2)}$ and $\chi^{(3)}$ non-linear interactions under continuous-wave or pulsed optical pumping \cite{JLu2020,Gaeta2019}. Finally, synthetic photonic lattices, realized as coupled optical waveguide arrays, enable the flexible engineering of evanescent coupling coefficients to arbitrary values by adjusting waveguide spacing and geometry \cite{Szameit2010}.
}

We emphasize that the model presented here is mathematically consistent and analytically solvable, providing insight into the behavior of coupled bosonic modes under symplectic transformations. Although it does not aim to describe the full complexity of a nonstationary cavity field with moving boundaries, it offers an analog for studying quantum correlations, entanglement, and squeezing in a controlled setting.
\medskip\\
{\bf Generators of nonstandard two-mode squeezed states.} According to formulas (3.1a), (3.1b), and (3.1c) on page 740 of Milburn \cite{Milburn1984}, the generators of the Lie algebra $\mathrm{Sp}\left(2n;\mathbb{R}\right)$ are given by the following $n(2n+1)$ bilinear operators
\begin{equation}\label{func:alg_gen}
	\begin{aligned}
		\hat{H}_i=&\frac{1}{2}\left(\hat{a}_i^\dagger \hat{a}_i+\hat{a}_i \hat{a}_i^\dagger  \right), \quad i=1,2,3,\dots,n;
		\\
		\hat{C}_{i,j}=&\hat{a}_i^\dagger\hat{a}_j, \quad i\neq j, \; i,j=1,2,3,\dots,n;
		\\
		\hat{a}_i^\dagger\hat{a}_j^\dagger, \quad   &
		\hat{a}_i\hat{a}_j,  \quad i\leq j, \, j=1,2,3,\dots,n.
	\end{aligned}
\end{equation}
When $n = 2$, there are ten generators given by the relations \eqref{func:alg_gen}. On the other hand, in the time-independent part of Eq.~\eqref{0020}, we have the generators
\begin{align}\label{T's}
\hat{T}_{4} &\equiv \mathrm{i}\left(\hat{a}_{1}\hat{a}_{2}^{\dagger} - \hat{a}_{1}^{\dagger}\hat{a}_{2}\right),\quad \hat{T}_{6} \equiv \mathrm{i}\left(\hat{a}_{1}^{\dagger}\hat{a}_{2}^{\dagger} - \hat{a}_{1}\hat{a}_{2}\right), 
\nonumber \\ 
\hat{T}_{8} &\equiv \frac{\mathrm{i}}{2}\left(\hat{a}_{1}^{\dagger}{}^{2} - \hat{a}_{1}^{2}\right),\qquad \hat{T}_{10} \equiv \frac{\mathrm{i}}{2}\left(\hat{a}_{2}^{\dagger}{}^{2} - \hat{a}_{2}^{2}\right),
\end{align}
where $\hat{T}_{4}$ belongs to the stationary subalgebra $\mathrm{u}(2)$, $\hat{T}_{8}$ and $\hat{T}_{10}$ belong to the Cartan subalgebra of $\mathrm{Sp}(4, \mathbb{R})$, and $\hat{T}_{6}$ is a non-standard element representing parametric amplification. In this case, the combination of the four elements in a Hamiltonian can be viewed as a photonic system generating two-mode squeezed states \cite{Milburn1984,Walls1983,Luis1995}. Thus, from the algebraic standpoint, we can think of the time-independent part of the effective factorized Hamiltonian \eqref{0020} as an optical realization of some device that gives rise to the generation of squeezed states of two modes. With this in mind, we can proceed with our analysis and interpret $v/c$ as a coupling parameter $g$ in a photonic system.
{\color{black} To explicitly justify this reinterpretation, note that the bilinear operator combinations in Eq.~\eqref{0020}, namely the beam-splitter-like mode conversion term $\propto (\hat{a}_1 \hat{a}_2^\dagger - \hat{a}_1^\dagger \hat{a}_2)$ and the parametric two-mode squeezing term $\propto (\hat{a}_1^\dagger \hat{a}_2^\dagger - \hat{a}_1 \hat{a}_2)$, are precisely the elements generating the Lie algebra $\mathrm{Sp}(4, \mathbb{R})$. In integrated photonic architectures, such as evanescently coupled optical waveguides or parametric optical devices, the evolution of quantum modes is governed by identical bilinear Hamiltonians; the difference is that, in those optical platforms, the dimensionless prefactors multiplying these algebraic generators are dictated by spatial overlap integrals, evanescent coupling coefficients, or non-linear optical susceptibilities ($\chi^{(2)}/\chi^{(3)}$) combined with pump field amplitudes, rather than by a physical mirror velocity $v$. Since the quantum dynamical evolution and the resulting non-classical correlations depend strictly on this Lie-algebraic structure and on the numerical strength of the prefactors, and not on their physical origin, the mathematical mapping $v/c \to g$ is fully rigorous. This equivalence allows us to decouple the model from the mechanical kinematic constraint $\vert{}v/c\vert{} < 1$ and treat $g$ as an effective coupling strength that can be engineered into the strong- or ultra-strong-interaction regime ($g > 1$) in photonic hardware.

To clarify the underlying mathematical framework of our diagonalization procedure, recall that the Lie algebra $\mathrm{sp}(4, \mathbb{R})$ admits non-equivalent Cartan decompositions associated with the maximal subalgebras $\mathrm{u}(2)$, $\mathrm{sp}(2, \mathbb{R}) \oplus \mathrm{sp}(2, \mathbb{R})$, and $\mathrm{sp}(2, \mathbb{C})$. In our scheme, the unitary transformations $\hat{R}_1$, $\hat{R}_2$, and $\hat{R}_3$ exploit the Cartan decomposition anchored on the maximal compact subalgebra $\mathrm{u}(2) \cong \mathrm{u}(1) \oplus \mathrm{su}(2)$ together with its non-compact subspace complement. Physically, the two pieces of this decomposition play distinct roles. Generators residing in the compact subalgebra $\mathrm{u}(2)$ preserve the total photon number and generate passive linear-optical transformations, physically realized via phase shifters and beam splitters, whereas the non-compact generators, spanning $\mathrm{sp}(2, \mathbb{R}) \oplus \mathrm{sp}(2, \mathbb{R})$, alter the total photon count and drive active non-linear processes such as single-mode squeezing, two-mode squeezing, and parametric amplification. Because the effective cavity Hamiltonian in Eq.~(2) contains both mode-conversion ($\mathrm{u}(2)$) and parametric-squeezing ($\mathrm{sp}(2,\mathbb{R})$) terms, complete diagonalization cannot be achieved through passive $\mathrm{u}(2)$ rotations alone; it strictly requires a structured sequence combining active squeezing transformations ($\hat{R}_1, \hat{R}_3$) with a passive beam-splitter rotation ($\hat{R}_2$).
}

\subsection{Diagonalization}
We choose to diagonalize this Hamiltonian in the space spanned by the \emph{position} and \emph{momentum} operators, since we can define a time-independent set of unitary transformations that uncouple the two modes in configuration space. For this, a selected map between $\{\hat{a}_{k}^{\dagger}, \hat{a}_{k}\}$ and $\{\hat{x}_{k}, \hat{p}_{k}\}$ is defined by the dimensionless relations
\begin{equation}
\hat{a}_{k} = \sqrt{\frac{k\pi}{2}}\left(\hat{x}_{k} + \frac{\mathrm{i}}{k\pi}\hat{p}_{k}\right),\qquad
\hat{a}_{k}^{\dagger} = \sqrt{\frac{k\pi}{2}}\left(\hat{x}_{k} - \frac{\mathrm{i}}{k\pi}\hat{p}_{k}\right), \qquad k=1,2,3, \dots,
\end{equation}
where the canonical commutation relation $[\hat{a}_{k},\hat{a}_{j}^{\dagger}] = \delta_{j,k}$ implies $[\hat{x}_{k}, \hat{p}_{j}] = \mathrm{i} \delta_{j,k}$. By definition, inverse operators $\left\{\hat{x}_{k},\hat{p}_{k}\right\}$ do not have associated dimensions. We can now map the terms in the time-independent Hamiltonian, resulting in
\begin{equation}
\begin{aligned}
    \hat{a}_{1}^{\dagger}\hat{a}_{1} &\mapsto \frac{1}{2}\left(\pi\hat{x}_{1}^{2} + \frac{\hat{p}_{1}^{2}}{\pi}\right) - \frac{1}{2}\hat{\mathbb{I}}_{1},\\
    \hat{a}_{2}^{\dagger}\hat{a}_{2} &\mapsto \frac{1}{2}\left(2\pi\hat{x}_{2}^{2} + \frac{\hat{p}_{2}^{2}}{2\pi}\right) - \frac{1}{2}\hat{\mathbb{I}}_{2},\\
    \hat{a}_{j}^{\dagger 2} - \hat{a}_{j}^{2} &\mapsto -\mathrm{i}\left(\hat{x}_{j}\hat{p}_{j} + \hat{p}_{j}\hat{x}_{j}\right),\\
    \hat{a}_{1}^{\dagger}\hat{a}_{2}^{\dagger} - \hat{a}_{1}\hat{a}_{2} &\mapsto -\sqrt{2}\mathrm{i}\left(\frac{\hat{x}_{1}\hat{p}_{2}}{2} + \hat{p}_{1}\hat{x}_{2}\right),\\
    \hat{a}_{1}\hat{a}_{2}^{\dagger} - \hat{a}_{1}^{\dagger}\hat{a}_{2} &\mapsto -\sqrt{{2}}\mathrm{i}\left(\frac{\hat{x}_{1}\hat{p}_{2}}{2} - \hat{p}_{1}\hat{x}_{2}\right),
\end{aligned}
\end{equation}
leading to a position-momentum version of the effective Hamiltonian \eqref{0020}
\begin{align}
\hat{H}_{\texttt{eff}}=&\frac{\hbar}{q_{0} + vt}\left\{\sum\limits_{k = 1}^{2}\left[\frac{c}{2}\left(k^{2}\pi^{2}\hat{x}_{k}^{2} + \hat{p}_{k}^{2}\right) + \frac{v}{4}\left(\hat{x}_{k}\hat{p}_{k}+\hat{p}_{k}\hat{x}_{k}\right) \right] 
\right. \nonumber \\   &   \left.
+ \frac{2v}{3}\left(\hat{x}_{1}\hat{p}_{2} - \hat{x}_{2}\hat{p}_{1}\right)-\frac{3\pi c}{2}\right\}.
\end{align}
We define the time-independent Hamiltonian as the part inside the braces,
\begin{equation}
	\hat{\mathcal{H}}_{xp}\equiv
	\hbar\sum\limits_{k = 1}^{2}\left[\frac{c}{2}\left(k^{2}\pi^{2}\hat{x}_{k}^{2} + \hat{p}_{k}^{2}\right) + \frac{v}{4}\left(\hat{x}_{k}\hat{p}_{k}+\hat{p}_{k}\hat{x}_{k}\right) \right] + \frac{2\hbar v}{3}\left(\hat{x}_{1}\hat{p}_{2} - \hat{x}_{2}\hat{p}_{1}\right).
\end{equation}
We can diagonalize $\hat{\mathcal{H}}_{xp}$ using a set of unitary transformations (unitary, as we require $\chi$ and $\xi$ to be real),
\begin{equation}\label{0070}
    \hat{R}_{1} = \exp\left[-\mathrm{i}\frac{v}{4c}\left(\hat{x}_{1}^{2} + \hat{x}_{2}^{2}\right)\right],\quad
    \hat{R}_{2} = \exp\left(-\mathrm{i}\chi\hat{x}_{1}\hat{x}_{2}\right),\quad
    \hat{R}_{3} = \exp\left(-\mathrm{i}\xi\hat{p}_{1}\hat{p}_{2}\right),
\end{equation}
acting on the position and momentum operators as
\begin{equation}
\begin{aligned}
    \hat{R}_{1}^{\dagger}\hat{p}_{1}\hat{R}_{1} = \hat{p}_{1} - \frac{v}{2c}\hat{x}_{1},\quad &\hat{R}_{1}^{\dagger}\hat{p}_{2}\hat{R}_{1} = \hat{p}_{2} - \frac{v}{2c}\hat{x}_{2},\\
    \hat{R}_{2}^{\dagger}\hat{p}_{1}\hat{R}_{2} = \hat{p}_{1} - \chi\hat{x}_{2},\quad &\hat{R}_{2}^{\dagger}\hat{p}_{2}\hat{R}_{2} = \hat{p}_{2} - \chi\hat{x}_{1},\\
    \hat{R}_{3}^{\dagger}\hat{x}_{1}\hat{R}_{3} = \hat{x}_{1} + \xi\hat{p}_{2},\quad &\hat{R}_{3}^{\dagger}\hat{x}_{2}\hat{R}_{3} = \hat{x}_{2} + \xi\hat{p}_{1}.
\end{aligned}
\end{equation}
{\color{black}
For notational simplicity, we define the composite unitary operator that performs the total transformation sequence as
\begin{equation}\label{eq0110} 
\hat{U} \equiv \hat{R}_1 \hat{R}_2 \hat{R}_3,
\end{equation}
so that $\hat{\mathcal{H}}_{R} = \hat{U}^\dagger \hat{H}_{xp} \hat{U}$.
}
Taking this last set of mappings, we can put it in $\hat{\mathcal{H}}_{xp}$ to arrive at a set of harmonic oscillators that are still coupled in the position-momentum terms,
\begin{equation}
\begin{aligned}\label{Hxp}
    \hat{\mathcal{H}}_{R} &= \hat{U}^\dagger \hat{\mathcal{H}}_{xp}  \hat{U}\\
    & = \hbar\left[\sum\limits_{k = 1}^{2}\left(\alpha_{k}\hat{x}_{k}^{2} + \beta_{k}\hat{p}_{k}^{2}\right) + \gamma_{12}\hat{x}_{1}\hat{p}_{2} + \gamma_{21}\hat{x}_{2}\hat{p}_{1}\right],
\end{aligned}
\end{equation}
where
\begin{equation}\label{0090}
\begin{aligned}
    \alpha_{1} &= -\frac{v^2}{8 c}+\frac{c \chi ^2}{2}+\frac{\pi ^2 c}{2}-\frac{2 v \chi }{3},\\
    \alpha_{2} &=-\frac{v^2}{8 c}+\frac{c \chi ^2}{2}+2 \pi ^2 c+\frac{2 v \chi }{3},\\
    \beta_{1} &= \frac{1}{2} c \xi ^2 \chi ^2+2 \pi ^2 c \xi ^2-c \xi  \chi -\frac{\xi ^2 v^2}{8 c}+\frac{c}{2}+\frac{2}{3} \xi ^2 v \chi -\frac{2 \xi  v}{3},\\
    \beta_{2} &=\frac{1}{2} c \xi ^2 \chi ^2+\frac{1}{2} \pi ^2 c \xi ^2-c \xi  \chi -\frac{\xi ^2 v^2}{8 c}+\frac{c}{2}-\frac{2}{3} \xi ^2 v \chi +\frac{2 \xi  v}{3},\\
    \gamma_{12} &=c \xi  \chi ^2+\pi ^2 c \xi -\frac{\xi  v^2}{4 c}-c \chi -\frac{4 \xi  v \chi }{3}+\frac{2 v}{3},\\
    \gamma_{21} &=c \xi  \chi ^2+4 \pi ^2 c \xi -\frac{\xi  v^2}{4 c}-c \chi +\frac{4 \xi  v \chi }{3}-\frac{2 v}{3}.
\end{aligned}
\end{equation}
We can remove these position-momentum couplings by setting their coefficients equal to zero, $\gamma_{12} = \gamma_{21} = 0$, which in turn gives a pair of values for $\chi$ and $\xi$,
\begin{equation}\label{0110}
    \chi_{\pm} =\pm \frac{\sqrt{\Delta }}{8 c^2 v}-\frac{9 \pi ^2 c}{8 v},\qquad
    \xi_{\pm} = \pm \frac{4c^2 v}{\sqrt{\Delta}},
\end{equation}
where we have defined
\begin{equation}
    \Delta = 81 \pi^{4}c^6 + 160 \pi^{2}c^4 v^{2}-16c^2 v^{4}.
\end{equation}
We will only consider the $+$ sign in expressions \eqref{0110} for $\chi$ and $\xi$,  since for $v=0$, the original Hamiltonian in \eqref{0020} is just the sum of two uncoupled quantum harmonic oscillators, and in that case, we want our transformations \eqref{0070} to be the identity.
For the transformations $\hat{R}_2$ and $\hat{R}_3$ to be unitary, $\chi$ and $\xi$ must be real, which requires $\Delta > 0$, implying that the validity range must be
\begin{equation}
-\frac{1}{2} \sqrt{\sqrt{481}+20} \; \pi \leq \frac{v}{c} \leq \frac{1}{2} \sqrt{\sqrt{481}+20} \; \pi,
\qquad \left( \left\vert\frac{v}{c}\right\vert \lesssim 10.1716 \right),
\end{equation}
where, for the usual cavity field, the physically acceptable values were within the open interval $v/c \in [0, 1)$. Now, as discussed in Section \ref{photonic}, we may consider this {\color{black}normalized velocity} to be a coupling strength $g$ in a two-mode squeezed-state photonic system generator. Thus, the range may symbolically lie in $\left\vert g\right\vert \lesssim 10.1716$. From now on, we refer to $g$ only.

\subsection{Diagonalized system}
We will continue to analyze the equations, keeping the mathematical range within that of a photonic system. The values of $\chi$ and $\xi$ are now defined, so we arrive at a set of time-independent uncoupled harmonic oscillators,
\begin{equation}\label{HR}
    \hat{\mathcal{H}}_{R} = \hbar\sum\limits_{k = 1}^{2}\left(\alpha_{k}\hat{x}_{k}^{2} + \beta_{k}\hat{p}_{k}^{2}\right),
\end{equation}
where the coefficients involved are explicitly given by
\begin{equation}\label{abcoeffs}
\begin{aligned}
    \alpha_{1} =\frac{1}{c^2}\left(\frac{\Delta }{64 g^2c^3 }\mp\sqrt{\Delta } \left(\frac{1}{12}+\frac{9 \pi ^2}{64 g^2}\right)\right),\quad&
    \alpha_{2} =\frac{1}{c^2}\left(\frac{\Delta }{64 c^3 g^2}\mp\sqrt{\Delta } \left(\frac{9 \pi ^2}{64 g^2}-\frac{1}{12}\right)\right),\\  
    \beta_{1} =\frac{c}{4}\mp\frac{c^4}{\sqrt{\Delta }}\left( \frac{4 g^2}{3}-\frac{9 \pi ^2}{4}\right),\quad&
    \beta_{2} =\frac{c}{4}\pm\frac{c^4}{\sqrt{\Delta }}\left( \frac{4 g^2}{3}+\frac{9 \pi ^2}{4}\right).
\end{aligned}
\end{equation}
From this set of relations, we can straightforwardly conclude that the value $g = 0$ is forbidden since there are denominators in $\alpha_{k}$ that involve it. Therefore, the photonic coupling $g$ must be different from zero at any time.

\begin{figure}[H]
	\centering
	\includegraphics[width = \textwidth, keepaspectratio]{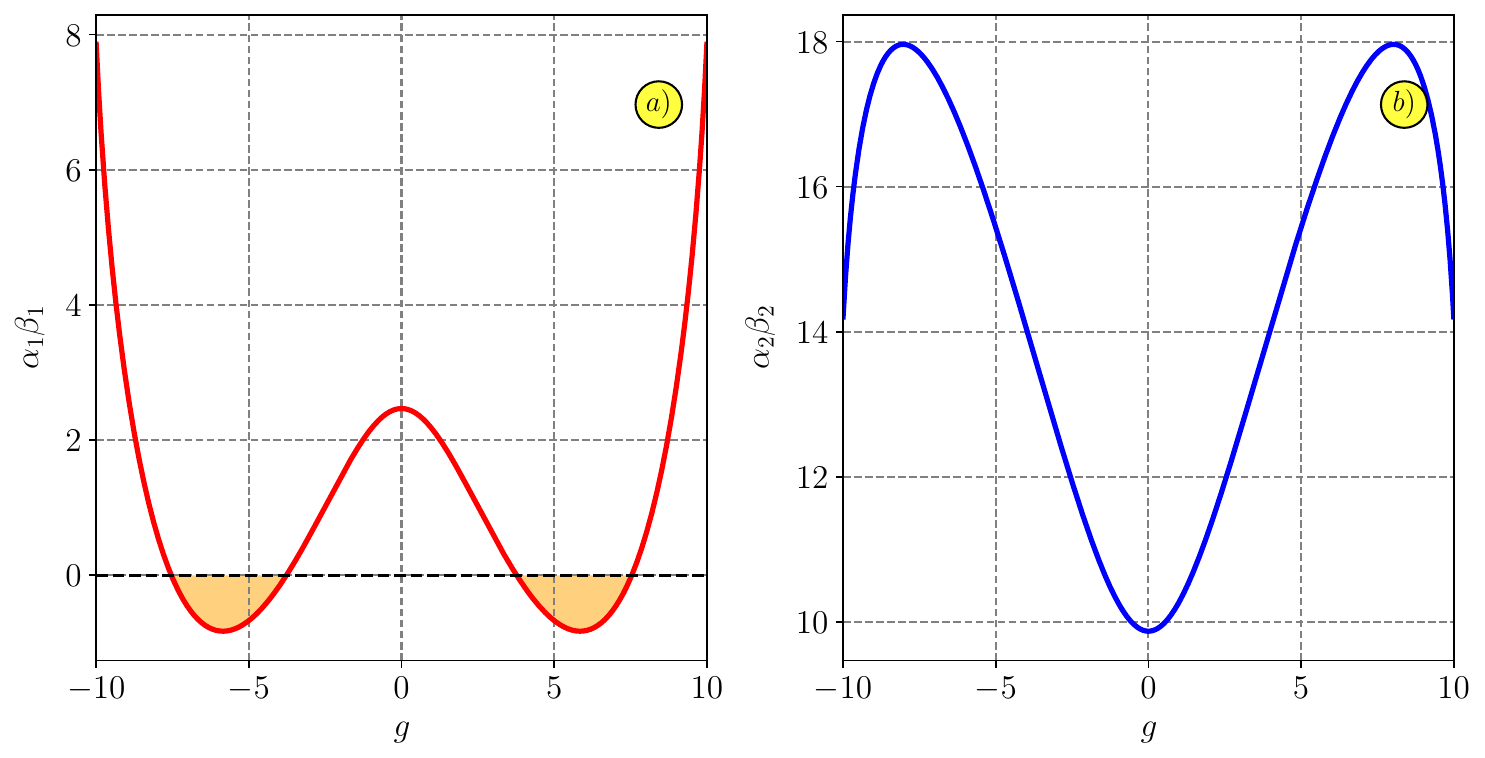}
	\caption{Products $\alpha_k\beta_k$ as functions of the effective coupling $g$ within the unitary range $|g|\lesssim 10.1716$ (plus sign chosen in Eq.~(17)). Panel a) $\alpha_1\beta_1$ versus $g$, showing a symmetric interval around the origin where the product becomes negative, leading to imaginary contributions to the pseudo-energy. Panel b) $\alpha_2\beta_2$ versus $g$, which remains positive over a wider range, highlighting the mode-dependent onset of forbidden (non--harmonic-oscillator) regions.}
	\label{fig:abprod}
\end{figure}

\begin{figure}[H]
    \centering
    \includegraphics[width = \textwidth, keepaspectratio]{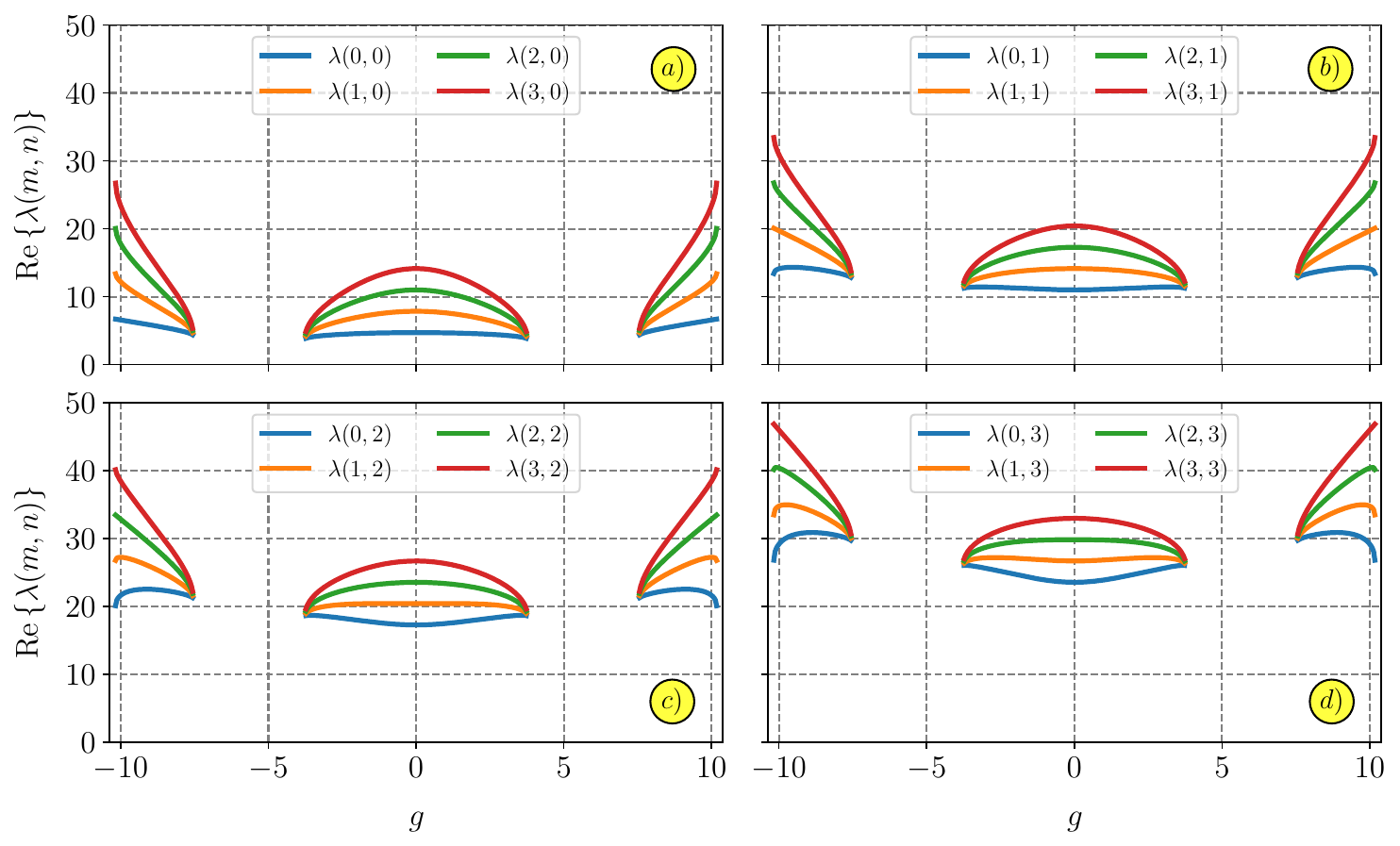}
    \caption{Real part of the pseudo-energy spectrum $\lambda(m,n)$ as a function of the effective coupling $g$ over the interval $|g|\leq 10.1716$. Continuous branches correspond to regimes where the products $\alpha_{k}\beta_{k}$ are positive. Here, the system admits a harmonic-oscillator interpretation, while gaps and truncated curves indicate intervals where $\alpha_{1}\beta_{1}<0$, resulting in complex eigenvalues and dynamically forbidden regions.  Panel a) Ground-manifold spectrum with $n=0$ and $m=0,1,2,3$, illustrating the onset of spectral gaps as $g$ increases.  Panel b) First excited manifold with $n=1$ and $m=0,1,2,3$, showing a similar structure shifted to higher energies and enhanced sensitivity to the forbidden intervals. Panel c) Second excited manifold with $n=2$ and $m=0,1,2,3$, where the widening of inaccessible regions becomes more pronounced. Panel d) Third excited manifold with $n=3$ and $m=0,1,2,3$, highlighting the progressive deformation and fragmentation of the spectrum at larger excitation numbers.}
    \label{fig:lambdas}
\end{figure}

The pseudoenergy spectrum of the Hamiltonian $\hat{\mathcal{H}}_{R}$ is the direct sum of the energies of two quantum harmonic oscillators with eigenvalues $m$ and $n$, defined as
\begin{equation}\label{lambdas}
\begin{aligned}
    \lambda(m, n) = 2\hbar\sqrt{\alpha_{1}\beta_{1}}\left(m + \frac{1}{2}\right) + 2\hbar\sqrt{\alpha_{2}\beta_{2}}\left(n + \frac{1}{2}\right),
\end{aligned}
\end{equation}
where $m,n\in\mathbb{N}$. It is worth analyzing the products inside the square roots, especially in the first oscillator. From now on until the other statement, we will work in normalized units with $\hbar = 1$. Due to the definitions of $\alpha_{k}$ and $\beta_{k}$, we need to choose one of the two signs involved; for this, we take the first sign in the relations \eqref{abcoeffs}, and then calculate the products $\alpha_{1}\beta_{1}$ and $\alpha_{2}\beta_{2}$ as functions of the parameter $g$. We can see, in Fig.~\ref{fig:abprod}, that there will be imaginary components because the product of the coefficients becomes negative at some subintervals (orange region).
This behavior will have consequences on the spectrum $\lambda(m,n)$, as Fig.~\ref{fig:lambdas} depicts. There we plot the first sixteen values of $\lambda(m,n)$, with combinations $m, n\in\{0, 1, 2, 3\}$. Regions where the products $\alpha_{1}\beta_{1}$ become negative appear as voids. We can therefore conclude that these values are inaccessible (forbidden), but they introduce a new kind of dynamic in the system, such as converting the oscillator into a repulsive one.

{\color{black} The value $g \approx 3.75$ serves as a critical threshold because it corresponds to the exact root at which the product coefficient vanishes, i.e., $\alpha_1(g)\beta_1(g) = 0$ (as illustrated in Figure~\ref{fig:abprod}a). In the diagonalized picture, the effective frequency of the first mode behaves as $\Omega_1 \propto \sqrt{\alpha_1(g)\beta_1(g)}$, so the sign of this product directly controls the physical character of the mode. Three consequences follow from this dependence. First, as $g$ approaches $3.75$ from below, the effective frequency softens to zero ($\Omega_1 \to 0$), flattening the effective confinement potential. Second, this softening marks a transition in phase space: for example $0 < g < 3.75$, the product $\alpha_1 \beta_1 > 0$ guarantees real effective frequencies and closed, bounded phase-space trajectories corresponding to stable harmonic oscillations, and at the critical point $g \approx 3.75$ itself, the potential transitions from an attractive parabolic well to a flat potential. Beyond this point, for $g > 3.75$, the product becomes strictly negative ($\alpha_1 \beta_1 < 0$), rendering the effective frequency purely imaginary ($\Omega_1 \in i\mathbb{R}$); physically, this converts the harmonic potential into an inverted (repulsive) potential, signaling a transition to a regime of parametric instability in which stationary bound states no longer exist and quantum field fluctuations grow exponentially. It is important to emphasize that this threshold, and the emergence of complex eigenvalues for $g > 3.75$, reflect a physical stability threshold rather than a mere mathematical artifact. Physically, $g$ quantifies the ratio between the cross-mode position-momentum interaction and the intrinsic restoring potentials of the individual modes. The regime $g > 3.75$ corresponds to an over-coupled parametric interaction in which cross-coupling dominates, triggering a dynamical instability (unbounded motion or runaway energy transfer) within the system. In this unstable regime, the linearized quadratic Hamiltonian approximation breaks down, requiring higher-order nonlinearities or dissipative channels to maintain physical steady states. Restricting the domain to $0 < g \lesssim 3.75$ therefore guarantees the dynamical stability of the linearized two-mode system and ensures real eigenvalues and unitary diagonalizability. This photonic-domain stability threshold should not be confused with the kinematic bound $|v/c| < 1$ of the original mechanical cavity: once $v/c$ is reinterpreted as the coupling strength $g$, that mechanical constraint no longer applies, and the relevant condition for physically stable, unitarily diagonalizable dynamics is instead set by $\alpha_1(g)\beta_1(g) > 0$, i.e., $g \approx 3.75$.
}

We can invert the set of transformations back and obtain the Hamiltonian $\hat{\mathcal{H}}_{R}$ in terms of new annihilation and creation operators, spawning a proper Fock space,
\begin{equation}
\begin{aligned}
    \hat{\mathcal{H}}_{R} &= \sum\limits_{k = 1}^{2}\left(\alpha_{k}\hat{x}_{k}^{2} + \beta_{k}\hat{p}_{k}^{2} - k\pi\hat{I}_{k}\right)\\
    &\rightarrow \sum\limits_{k = 1}^{2}\left(\sqrt{\alpha_{k}}\hat{x}_{k} - \mathrm{i}\sqrt{\beta_{k}}\hat{p}_{k}\right)\left(\sqrt{\alpha_{k}}\hat{x}_{k} + \mathrm{i}\sqrt{\beta_{k}}\hat{p}_{k}\right) - k\pi\hat{I}_{k}\\
    &\propto \sum\limits_{k = 1}^{2}\left(\alpha_{k}\hat{x}_{k}^{2} + \beta_{k}\hat{p}_{k}^{2}\right)\\ 
    &= \sum\limits_{k = 1}^{2}\left[\hat{A}_{k}^{\dagger}\hat{A}_{k} + \left(\sqrt{\alpha_{k}\beta_{k}} - k\pi\right)\hat{I}_{k}\right]
\end{aligned}
\end{equation}
where the new ladder operators are defined by
\begin{equation}
    \hat{A}_{k} = \sqrt{\alpha_{k}}\hat{x}_{k} + \mathrm{i}\sqrt{\beta_{k}}\hat{p}_{k},\qquad 
    \hat{A}_{k}^{\dagger} = \sqrt{\alpha_{k}}\hat{x}_{k} - \mathrm{i}\sqrt{\beta_{k}}\hat{p}_{k}.
\end{equation}
This explicitly demonstrates that a diagonalized Hamiltonian in the $\left\{\hat{x}, \hat{p}\right\}$ representation is isomorphically mapped to a diagonalized Hamiltonian in the $\left\{\hat{a}^{\dagger}, \hat{a}\right\}$ representation. However, when $\alpha_k$ and $\beta_k$ are negative, the mapping breaks down. In such cases, it is not possible to revert to the creation and annihilation operators ($\hat{a}, \hat{a}^\dagger$), rendering the isomorphism invalid. This breakdown significantly changes the system's dynamics, so the originally intended physical interpretation may no longer hold.

\section{Expectation values}\label{expec_valu}
So far, we have diagonalized the original Hamiltonian and identified the conditions under which the mapping is unitary. Using \eqref{HR} and the inverse of \eqref{Hxp}, we can calculate the expectation values. Although the inverse transformation is cumbersome, we can obtain it explicitly, allowing us to start calculating expectation values, as shown in the next section.

\subsection{Mapping between representations}
The expectation value of the photon number operator $\hat{n}_{k}\equiv\hat{a}^{\dagger}_{k}\hat{a}_{k}$ in mode $k$ can be carried on as
{\color{black}
\begin{align}
\hat{U} e^{\mi F(t) \hat{H}_R} \hat{U}^\dagger \left[ \hat{a}_k^\dagger \hat{a}_k \right] \hat{U} e^{-\mi F(t) \hat{H}_R} \hat{U}^\dagger =& \left( \hat{U} e^{\mi F(t) \hat{H}_R} \hat{U}^\dagger \hat{a}_k^\dagger \hat{U} e^{-\mi F(t) \hat{H}_R} \hat{U}^\dagger \right) 
\nonumber \\ & \times
\left( \hat{U} e^{\mi F(t) \hat{H}_R} \hat{U}^\dagger \hat{a}_k \hat{U} e^{-\mi F(t) \hat{H}_R} \hat{U}^\dagger \right),
\end{align}
}
where $F(t) = \int_{0}^{t}(q_{0} + v t')^{-1}dt'$, or any factorized function of time.

Since the transformation operators $\hat{R}_{j}$ given by Eq.~\eqref{0070} are defined to act on elements in the position-momentum $\{\hat{x}, \hat{p}\}$ space, we can apply the transformation using the map,
\begin{equation}
    \hat{a}_{k}^{\dagger} = \sqrt{\frac{k\pi}{2}}\left(\hat{x}_{k}-\frac{\mathrm{i}}{k\pi}\hat{p}_{k}\right),\qquad \hat{a}_{k}\equiv\left(\hat{a}_{k}^{\dagger}\right)^{\dagger},\qquad k\in\left\{1, 2\right\}.
\end{equation}
As we can see, we need to know two things: $i)$ how the position and momentum operators are transformed in the $\hat{R}_{j}^{\dagger}\left(\cdot\right)\hat{R}_{j}$ and $\hat{R}_{j}\left(\cdot\right)\hat{R}_{j}^{\dagger}$ order, which is, {\color{black}using the composite representation $\hat{U}$, can be written compactly as $\hat{x}'_k = \hat{U}^\dagger \hat{x}_k \hat{U}$ and $\hat{p}'_k = \hat{U}^\dagger \hat{p}_k \hat{U}$ (as well as their inverse transformations $\hat{U} \hat{x}_k \hat{U}^\dagger$ and $\hat{U} \hat{p}_k \hat{U}^\dagger$). The explicit expanded expressions for all eight transformation rules of the position and momentum operators can be found in the subappendix \ref{apAsub1}.}\\
$ii)$ How the propagator $e^{\mathrm{i}F(t)\hat{\mathcal{H}}_{R}}$ acts on the position and momentum operators, yielding
\begin{equation}
\begin{aligned}
e^{\mi F(t)\hat{\mathcal{H}}_{R}}\hat{x}_{j}e^{-\mi F(t)\hat{\mathcal{H}}_{R}} &= \hat{x}_{j}\cos\left(2\sqrt{\alpha_{j}\beta_{j}} F(t)\right) + \sqrt{\frac{\beta_{j}}{\alpha_{j}}}\;\hat{p}_{j}\sin\left(2\sqrt{\alpha_{j}\beta_{j}}F(t)\right),\\
e^{\mi F(t)\hat{\mathcal{H}}_{R}}\hat{p}_{j}e^{-\mi F(t)\hat{\mathcal{H}}_{R}} &= \hat{p}_{j}\cos\left(2\sqrt{\alpha_{j}\beta_{j}}F(t)\right) - \sqrt{\frac{\alpha_{j}}{\beta_{j}}}\;\hat{x}_{j}\sin\left(2\sqrt{\alpha_{j}\beta_{j}}F(t)\right),\; j\in\left\{1, 2\right\}.
\end{aligned}
\end{equation}
By knowing these two rules, we can perform the transformation of $\hat{x}_{k}$ and $\hat{p}_{k}$ accordingly, and then write the number operator with the inverse map $\{\hat{x}_{k}, \hat{p}_{k}\} \mapsto \{\hat{a}_{k}^{\dagger}, \hat{a}_{k}\}$.

\subsection{Photon generation and mode population dynamics}
From the relations found in the last section, we observe that the transformed creation-annihilation operators are linear combinations of operators of the two modes with coefficients $\eta_{l,m}$ obeying conjugation relations among them. See Appendix \ref{nexp_coeffs_1} for the detailed expressions. 
\begin{equation}\label{eq0250}
\begin{aligned}
\hat{U}e^{\mi F(t)\hat{\mathcal{H}}_{R}}\hat{U}^{\dagger}\left[\hat{a}_{1}\vphantom{\hat{a}_{1}^{\dagger}}\right]\hat{U}&e^{-\mi F(t)\hat{\mathcal{H}}_{R}}\hat{U}^{\dagger} =\\ &\eta_{1,1}\hat{a}_{1} + \eta_{1,2}\hat{a}_{1}^{\dagger} + \eta_{1,3}\hat{a}_{2} + \eta_{1,4}\hat{a}_{2}^{\dagger},
\end{aligned}
\end{equation}
\begin{equation}\label{eq0260}
\begin{aligned}
\hat{U}e^{\mi F(t)\hat{\mathcal{H}}_{R}}\hat{U}^{\dagger}\left[\hat{a}_{1}^{\dagger}\right]\hat{U}&e^{-\mi F(t)\hat{\mathcal{H}}_{R}}\hat{U}^{\dagger} =\\ &\eta_{3,1}\hat{a}_{1} + \eta_{3,2}\hat{a}_{1}^{\dagger} + \eta_{3,3}\hat{a}_{2} + \eta_{3,4}\hat{a}_{2}^{\dagger},
\end{aligned}
\end{equation}
where we have used identities relating coefficients that are complex conjugates of each other.
\begin{equation}\label{eq0270}
    \eta_{3, 1} = \eta_{1, 2}^{*},\qquad \eta_{3, 2} = \eta_{1, 1}^{*},\qquad \eta_{3, 3} = \eta_{1, 4}^{*},\qquad \eta_{3, 4} = \eta_{1, 3}^{*}.
\end{equation}

Similar to the calculations for the first mode, the resulting transformed creation and annihilation operators are combinations of the two-mode operators, with coefficients $\eta_{l, m}$ that satisfy conjugation relations among them. See Appendix \ref{nexp_coeffs_2} for the explicitly detailed expressions. 
\begin{equation}
\begin{aligned}
\hat{U}e^{\mi F(t)\hat{\mathcal{H}}_{R}}\hat{U}^{\dagger}\left[\hat{a}_{2}\vphantom{\hat{a}_{1}^{\dagger}}\right]\hat{U}e^{-\mi F(t)\hat{\mathcal{H}}_{R}}\hat{U}^{\dagger} &= \eta_{2,1}\hat{a}_{1} 
\\ &
+ \eta_{2,2}\hat{a}_{1}^{\dagger} + \eta_{2,3}\hat{a}_{2} + \eta_{2,4}\hat{a}_{2}^{\dagger},\\ \hat{U}e^{\mi F(t)\hat{\mathcal{H}}_{R}}\hat{U}^{\dagger}\left[\hat{a}_{2}^{\dagger}\right]\hat{U}e^{-\mi F(t)\hat{\mathcal{H}}_{R}}\hat{U}^{\dagger} &= \eta_{4,1}\hat{a}_{1} 
\\ &
+ \eta_{4,2}\hat{a}_{1}^{\dagger} + \eta_{4,3}\hat{a}_{2} + \eta_{4,4}\hat{a}_{2}^{\dagger},
\end{aligned}
\end{equation}
where we have used the identities between coefficients that are related by their own complex conjugates.
\begin{equation}
    \eta_{4, 1} = \eta_{2, 2}^{*},\qquad \eta_{4, 2} = \eta_{2, 1}^{*},\qquad \eta_{4, 3} = \eta_{2, 4}^{*},\qquad \eta_{4, 4} = \eta_{2, 3}^{*}.
\end{equation}

We can obtain the average number of photons in the first mode for the initial condition $|n,m\rangle$ as follows,
\begin{equation}
\begin{aligned}\label{n1}
    &\braket{\hat{a}_{1}^{\dagger}\hat{a}_{1}} = \langle m, n\vert \hat{U}e^{\mi F(t)\hat{\mathcal{H}}_{R}}\hat{U}^{\dagger}\left[\hat{a}_{1}^{\dagger}\hat{a}_{1}\right]\hat{U}e^{-\mi F(t)\hat{\mathcal{H}}_{R}}\hat{U}^{\dagger} \vert n, m \rangle\\
    &= \langle m, n\vert \left(\eta_{3,1}\hat{a}_{1} + \eta_{3,2}\hat{a}_{1}^{\dagger} + \eta_{3,3}\hat{a}_{2} + \eta_{3,4}\hat{a}_{2}^{\dagger}\right)\left(\eta_{1,1}\hat{a}_{1} + \eta_{1,2}\hat{a}_{1}^{\dagger} + \eta_{1,3}\hat{a}_{2} + \eta_{1,4}\hat{a}_{2}^{\dagger}\right) \vert n, m \rangle\\
    &= n\left(\eta_{1,2}\eta_{3,1} + \eta_{1,1}\eta_{3,2}\right) + m \left(\eta_{1,4}\eta_{3,3} + \eta_{1,3}\eta_{3,4}\right) + \eta_{1,2}\eta_{3,1} + \eta_{1,4}\eta_{3,3}
\end{aligned}
\end{equation}
and, proceeding in a similar form, we obtain the average number of photons in the second mode, we have
\begin{equation}
\begin{aligned}\label{n2}
    &\braket{\hat{a}_{2}^{\dagger}\hat{a}_{2}} = \langle m, n\vert \hat{U}e^{\mi F(t)\hat{\mathcal{H}}_{R}}\hat{U}^{\dagger}\left[\hat{a}_{2}^{\dagger}\hat{a}_{2}\right]\hat{U}e^{-\mi F(t)\hat{\mathcal{H}}_{R}}\hat{U}^{\dagger} \vert n, m \rangle\\
    &= \langle m, n\vert \left(\eta_{4,1}\hat{a}_{1} + \eta_{4,2}\hat{a}_{1}^{\dagger} + \eta_{4,3}\hat{a}_{2} + \eta_{4,4}\hat{a}_{2}^{\dagger}\right)\left(\eta_{2,1}\hat{a}_{1} + \eta_{2,2}\hat{a}_{1}^{\dagger} + \eta_{2,3}\hat{a}_{2} + \eta_{2,4}\hat{a}_{2}^{\dagger}\right) \vert n, m \rangle\\
    &= n\left(\eta_{2,2}\eta_{4,1} + \eta_{2,1}\eta_{4,2}\right) + m \left(\eta_{2,4}\eta_{4,3} + \eta_{2,3}\eta_{4,4}\right) + \eta_{2,2}\eta_{4,1} + \eta_{2,4}\eta_{4,3}.
\end{aligned}
\end{equation}

{\color{black} From the spectral and parameter analysis in Figures~\ref{fig:abprod} and \ref{fig:lambdas}, it is possible to identify the distinct operational ranges of $g$ that dictate the physical behavior of the system. Although the strict mathematical bound for unitary diagonalizing transformations extends up to $\vert{}g\vert{} \lesssim 10.1716$, Figure~1a shows that the product $\alpha_1 \beta_1$ remains strictly positive only within the continuous subinterval $0 < \vert{}g\vert{} \lesssim 3.75$. Beyond $g \approx 3.75$, $\alpha_1 \beta_1$ turns negative, giving rise to complex eigenenergies and spectral voids where the harmonic-oscillator picture breaks down. The well-behaved regime in which the system behaves as a pair of stable, coupled quantum harmonic oscillators is therefore bounded by $0 < g \lesssim 3.75$. Within this physically accessible domain, we specifically choose $g = 2.1$, $3.1$, and $3.5$ to illustrate three qualitatively different coupling regimes. The value $g = 2.1$ represents a moderate coupling regime, where photon creation occurs smoothly and the field statistics remain stable. The value $g = 3.1$ corresponds to a strong coupling regime, characterized by significantly enhanced two-mode squeezing and intermode entanglement. Finally, $g = 3.5$ captures a near-threshold strong coupling regime close to the critical value $g \approx 3.75$; in this region, the effective frequency flattens ($\alpha_1 \beta_1 \to 0$), triggering a steep increase in photon generation, rapid linear entropy growth, and non-classical statistical transitions in phase space just before entering the forbidden complex gap.

Equations \eqref{n1} and \eqref{n2} provide the expectation values of the photon number operators $\hat{n}_1$ and $\hat{n}_2$ for each mode. Applying the time evolution operator yields the time evolution of these expectation values for two initial conditions, $\ket{\psi(0)}\in\left\{\ket{0,0}, \ket{1,0}\right\}$. Fig.~\ref{fig:n1n2_gs} illustrates the time evolution for each mode. When both modes begin in the vacuum state, the squeezing of the two modes is evident and strongly depends on the coupling constant $g$. Near the critical value $g\approx 3.75$, the average number of photons increases rapidly. If one mode starts in the vacuum state and the other is excited, the squeezing effect is enhanced, as shown in the second row of the figure. This effect is further amplified when one mode has non-zero initial excitations. For comparison and validation, a simulation was performed using Julia's QuantumOptics.jl package \cite{Julia-2017,kramer2018quantumoptics}.

In the literature, most studies have focused on finding the expectation values of the photon number operator, as we have done up to this point. While this quantity clearly indicates particle creation and mode-population growth, it does not, by itself, fully characterize the quantum states generated by the dynamics. Distinct quantum states may exhibit identical mean photon numbers while differing substantially in their fluctuation properties, correlations, and degrees of coherence. In the present system, the underlying symplectic structure of the effective Hamiltonian naturally leads to nonclassical states whose characterization requires access to higher-order statistical information beyond $\langle \hat n_k \rangle$. In particular, quantities sensitive to number fluctuations, purity, and intermode correlations are necessary to distinguish between classical amplification, genuine squeezing, and entanglement. This becomes especially relevant near the transition regions of the effective coupling $g$, where qualitative changes in the spectrum and dynamics may occur even when the mean number of photons remains finite. For these reasons, a comprehensive understanding of the quantum evolution requires going beyond photon generation and examining appropriate statistical indicators, which we address in the following section.

\begin{figure}[H]
	\centering
	\includegraphics[width = \textwidth]{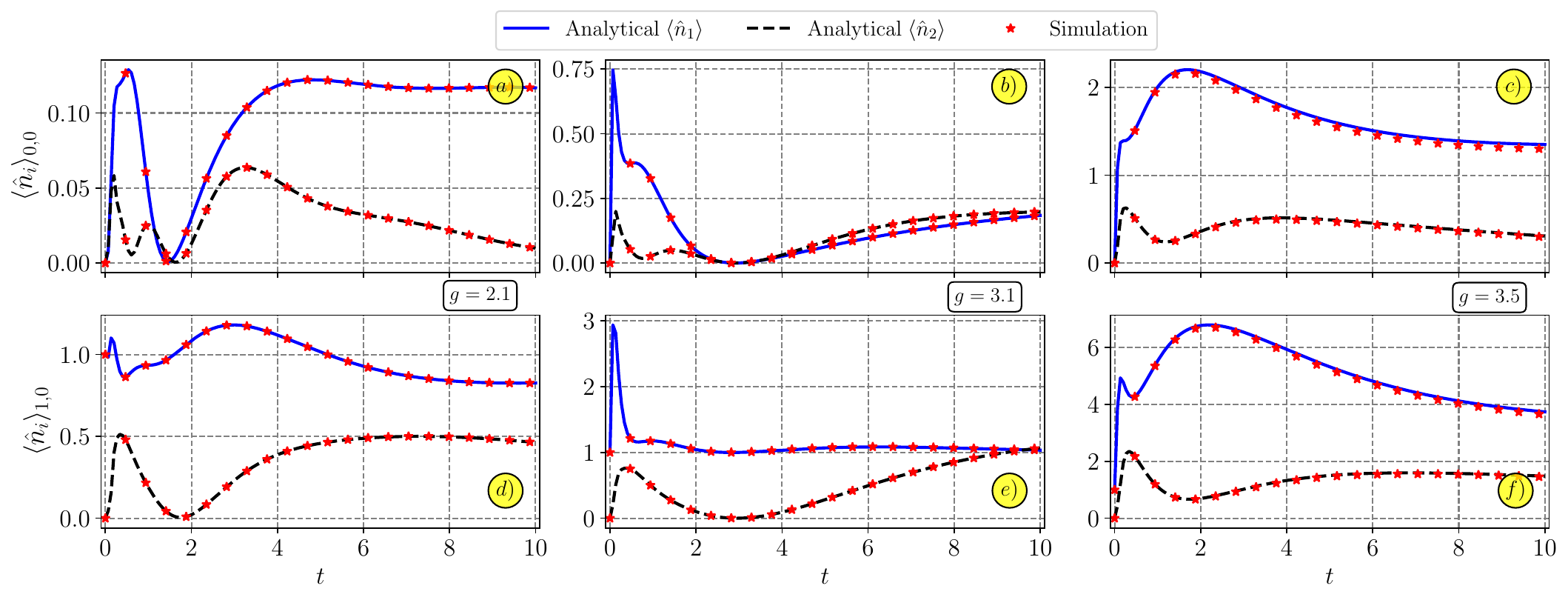}
	\caption{Time evolution of the photon-number expectation values $\langle \hat n_k\rangle$ for the two modes, obtained analytically from Eqs.~(27) and~(31) and compared with numerical simulations for different values of the effective coupling $g$. Solid curves represent the analytical predictions, while stars correspond to numerical results. Panels a) -- c) show the evolution from the initial vacuum state $|\psi(0)\rangle = |0,0\rangle$, whereas panels d) -- f) correspond to the initial state $|\psi(0)\rangle = |1,0\rangle$, with the first mode initially excited and the second remaining in vacuum. For each value of the parameter $g$, panels (a,d), (b,e), and (c,f) display the dynamics of mode~1 and mode~2, respectively, illustrating how initial excitation and increasing coupling enhance photon generation in both modes.}
	\label{fig:n1n2_gs}
\end{figure}

\section{Statistical signatures of nonclassicality and entanglement}\label{mix_entang}
To provide a more comprehensive description of the system's dynamics, we analyze statistical properties using the Mandel parameter, linear entropy, and the Wigner quasi-probability distribution function. This approach helps quantify the photon distribution and the degree of mixedness and entanglement in the system.

\subsection{Mandel parameter}
\begin{figure}[H]
	\centering
	\includegraphics[width = \textwidth]{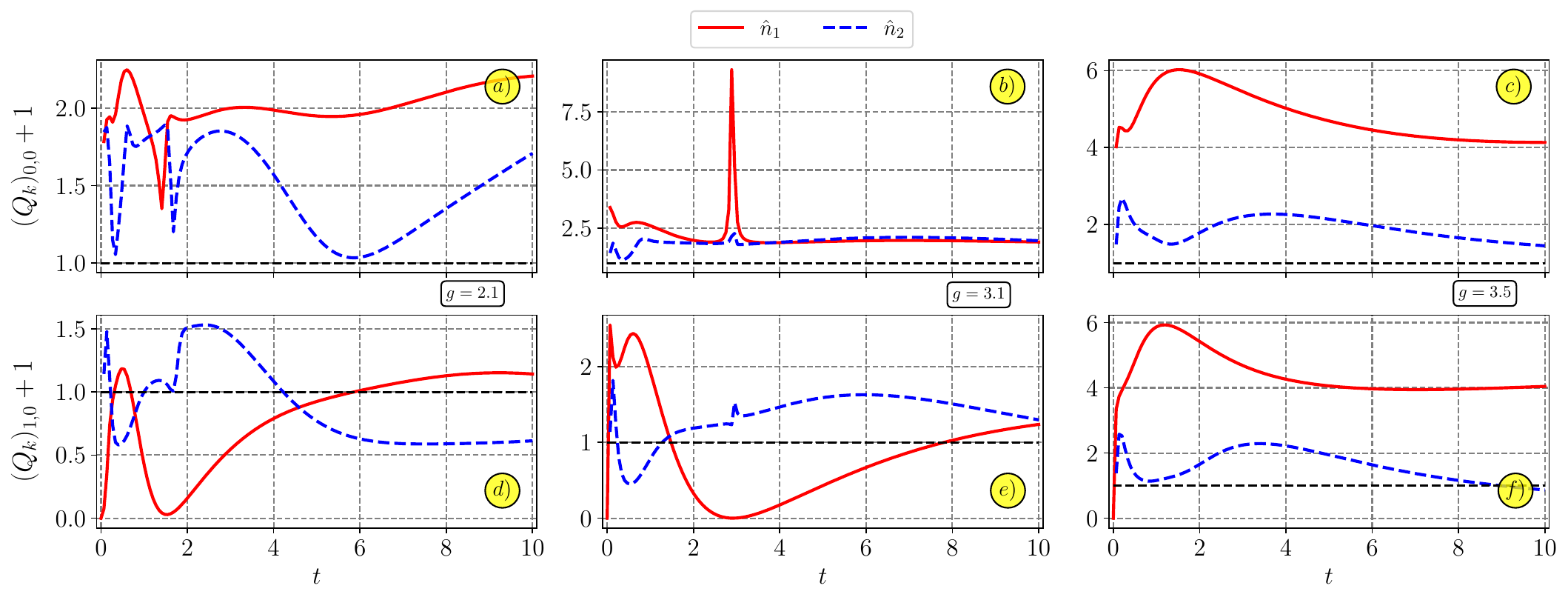}
	\caption{Time evolution of the Mandel parameter $Q_{k} + 1$ for the two modes, characterizing the photon-number statistics for different initial conditions and values of the effective coupling $g$. Values below unity indicate sub-Poissonian statistics, values equal to unity correspond to Poissonian statistics, and values above unity signal super-Poissonian behavior. Panels a) -- c) correspond to the initial vacuum state $|\psi(0)\rangle = |0,0\rangle$, while panels d) -- f) correspond to the initial state $|\psi(0)\rangle = |1,0\rangle$, with the first mode initially excited and the second in vacuum. For each value of $g$, panels (a, d), (b, e), and (c, f) show the Mandel parameter for mode~1 and mode~2, respectively, revealing the transition between non-classical and super-Poissonian statistics as the coupling strength and initial excitation are varied.}
	\label{fig:mandel_n1n2}
\end{figure}

We obtain valuable insights into the system's statistical behavior through the Mandel parameter \cite{Mandel1979}. Here, we use the shifted version of the parameter, defined as
\begin{equation}\label{Mandel}
	Q_{k} + 1 = \frac{\braket{\hat{n}_{k}^{2}} - \braket{\hat{n}_{k}}^{2}}{\braket{\hat{n}_{k}}},\qquad k = 1, 2,
\end{equation}
where for $Q_{k} + 1 < 1$ we have sub-Poissonian statistics, a signature of non-classical behavior whose limit is Fock states, and for $Q_{k} + 1 > 1$ we have super-Poissonian statistics similar to thermal distributions, whereas for $Q_{k} + 1 = 1$, it indicates Poissonian statistics, corresponding to the field coherent states. It is worth noticing that the right side of the equation \eqref{Mandel} is the normalized photon-number variance, commonly known as the Fano factor $F_{k}$, which quantifies deviations of the photon-number fluctuations from Poissonian statistics.

To determine the influence of the coupling parameter $g$ on the evolution of the photon statistics, we consider two representative initial conditions: a) both modes are initially prepared in the vacuum state, and b) one mode is prepared in the vacuum state while the other contains a single excitation. Figure~\ref{fig:mandel_n1n2} shows the corresponding time evolution of the parameter in \eqref{Mandel}. For case a), when $2 \le g \le 3$, the parameter $Q_{k} + 1$ remains above unity for both modes during most of the evolution, indicating super-Poissonian statistics. In contrast, for case b), the parameter exhibits intervals both above and below unity, demonstrating transitions between super-Poissonian and sub-Poissonian regimes. Increasing the coupling strength to $g = 3.5$ produces predominantly super-Poissonian statistics in the vacuum initial state. For the initially excited configuration, the evolution is also largely super-Poissonian, although the mode that starts from vacuum exhibits an initial interval with $Q_{k} + 1 < 1$, revealing a transient nonclassical regime characterized by sub-Poissonian fluctuations. These results strongly suggest that the effective coupling controls not only photon generation but also fluctuation properties, allowing the system to access different statistical regimes depending on both the initial state and the interaction strength.

\subsection{Linear entropy of entanglement}
\begin{figure}[H]
	\centering
	\includegraphics[width = \textwidth]{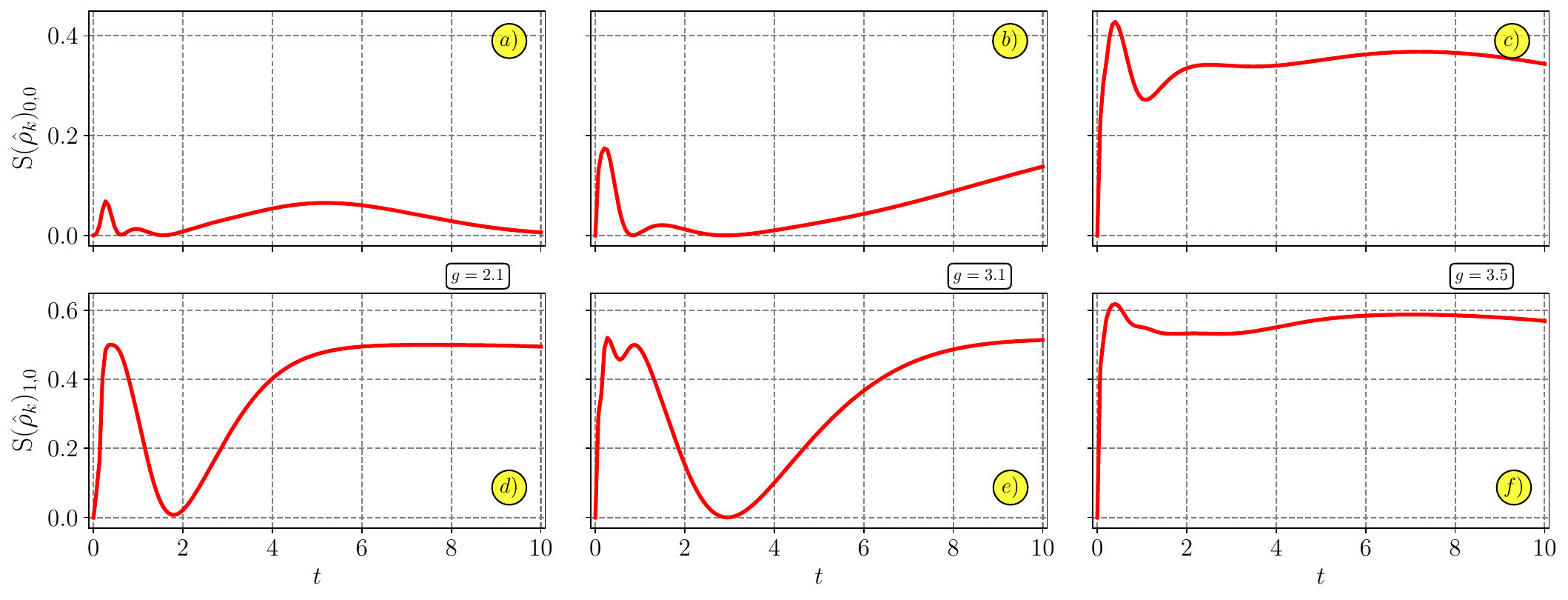}
	\caption{Time evolution of the linear entropy $S(\hat{\rho}_k)$ for the two modes, quantifying the degree of mixedness and entanglement generated during the dynamics for different initial conditions and values of the effective coupling $g$. Panels a) -- c) correspond to the initial vacuum state $|\psi(0)\rangle = |0,0\rangle$, while panels d) -- f) correspond to the initial state $|\psi(0)\rangle = |1,0\rangle$, with the first mode initially excited and the second in vacuum. For each value of $g$, panels (a,d), (b,e), and (c,f) show the linear entropy of mode~1 and mode~2, respectively, illustrating the growth of mixedness from initially pure states and its dependence on the coupling strength and initial excitation.}
	\label{fig:linen}
\end{figure}

{\color{black}
In this section, we analyze the linear entropy of the single-mode reduced density matrices as a function of time for several values of the effective coupling parameter $g$. Since the initial states under consideration ($\vert{}\psi(0)\rangle \in \{\vert{}0,0\rangle, \vert{}1,0\rangle\}$) are pure and the time evolution governed by the effective Hamiltonian is strictly unitary, the global bipartite state $\hat{\rho}_{1,2}(t) = \vert{}\psi(t)\rangle\langle\psi(t)\vert{}$ remains globally pure at all times, satisfying $\text{Tr}\left(\hat{\rho}_{1,2}^2\right) = 1$. Consequently, any emergence of mixedness in the single-mode reduced subsystems $\hat{\rho}_k = \text{Tr}_j \{\hat{\rho}_{1,2}\}$ ($k \neq j \in \{1,2\}$) obtained via partial tracing is a direct signature and quantitative measure of bipartite intermode entanglement. Formally, the linear entropy of the reduced state for mode $k$ is defined as \cite{Barnett2009, JuarezAmaro2020}
}
\begin{equation}\label{linen}
\begin{aligned}
	\mathrm{S}(\hat{\rho}_{k}) &:= \braket{\left(1 - {\hat\rho_{k}}\right)}\\
	&= 1 - \mathrm{Tr}\left\{{\hat\rho_{k}}^{2}\right\},
\end{aligned}
\end{equation}
where $\mathrm{Tr}\left\{{\hat\rho_{k}}^{2}\right\}$ is a quantity called purity, having the bounds
\begin{equation}
	\mathrm{Tr}\left\{{\hat\rho_{k}}^{2}\right\} = \sum\limits_{n}(\rho_{k})_{n}^{2} \leq \sum\limits_{n} (\rho_{k})_{n} = 1.
\end{equation}
For pure states, the linear entropy is zero; for mixed states, it is between zero and one, depending on the degree of mixing. Furthermore, linear entropy is related to von Neumann entropy as the lower bound $\mathrm{S}(\hat{\rho}_{k}) \leq -\mathrm{Tr}\left\{\hat{\rho}_{k}\ln \hat{\rho}_{k}\right\}$.

In Fig.~\ref{fig:linen}, we show the linear entropy \eqref{linen} for the states previously studied. As shown, it starts at zero in all panels, indicating that the system is prepared in a pure state. As time progresses, the linear entropy increases non-monotonically, indicating greater mixing among the modes. For the vacuum state and $g=2.1$, we observe a small amount of mixing at all times, and for sufficiently long times, the system tends toward a pure state. Furthermore, for the vacuum state and larger coupling $g$, the degree of mixing increases with time. In the case where there is an excitation in the initial state, the linear entropy starts at zero, and irrespective of the value taken by $g$, it increases rapidly and then shows a decrease whose magnitude depends on the value of $g$ and attains an asymptotic value for larger times.

\subsection{Wigner function}
For a bipartite system, the Wigner function of each subsystem is defined by the expansion displaced on the basis of the quantum harmonic oscillator \cite{MoyaBook2011,SchleichBook2001}, given by
\begin{equation}\label{wfunc}
	W_k(\alpha)= \frac{1}{\pi^{2}}\int\,d^{2}\beta\exp\left(\alpha\beta^{*} - \alpha^{*}\beta\right)\mathrm{Tr}\left\{\hat{\rho}_{k}\hat{D}\left(\beta\right)\right\},
\end{equation}
where $\hat{D}(\beta) = \exp\left(\beta\hat{a}^{\dagger} - \beta^{*}\hat{a}\right)$ is the Glauber displacement operator. Given that we know the density matrix for an initial state $\ket{\psi_{0}}$, under the time evolution $\hat{U}(t)\ket{\psi_{0}} = \exp\left(-\mathrm{i} t \hat{H}\right)\ket{\hat{\psi}_{0}} \equiv \hat{U}\exp\left(-\mathrm{i}F(t)\hat{\mathcal{H}}_{R}\right)\hat{U}^{\dagger}\ket{\psi_{0}}$, the Wigner function can be easily obtained for every time $t$.

In Fig.~\ref{fig:wfunc}, we show the evolution in phase space of the Wigner function for the first mode $W_1(\alpha)$, for selected initial states and pairs $\{t, g\}$. The first two rows, a)-b), show initial Gaussian profiles, given that there are no excitations in each mode. They evolve continuously as all-positive squeezed states. On the other hand, in the next two rows, c)-d), we see that the observed mode has a negative value because the initial state has one excitation. Again, the state evolves continuously as a squeezed state, preserving the negativity. Interestingly, there exist combinations $\{t_{3}, g = 3.5\}$ in row d) where the squeezed states become all-positive, as if they originated from a null-excitation initial state. According to the standard interpretation of the Wigner function, the presence of negative regions constitutes a sufficient signature of nonclassicality, while a strictly positive distribution allows for an effective classical description, though it does not imply classicality in a fundamental sense. In this light, the disappearance of negativity observed for specific combinations of time and strong coupling (e.g., $g = 3.5$) should be understood as a qualitative modification of the reduced-state description rather than as a loss of quantum coherence. This behavior occurs near the coupling-induced transition identified in the diagonalized spectrum, where the effective dynamics departs from that of bounded harmonic oscillators and acquires features analogous to repulsive or non–energy-bounded quadratic systems. In such regimes, the phase-space evolution generated by the underlying $\mathrm{Sp}(4,\mathbb{R})$ transformations can initially drive nonclassical states into configurations whose Wigner functions become entirely positive, resembling classical squeezed or thermal-like distributions despite their quantum origin. This observation reinforces the interpretation that effective coupling $g$ controls not only photon generation but also the qualitative structure of the dynamical generators, thereby shaping the observable signatures of nonclassicality and delimiting the regimes where a harmonic-oscillator picture remains meaningful.

\begin{figure}[H]
 	\centering
 	\includegraphics[width = \textwidth]{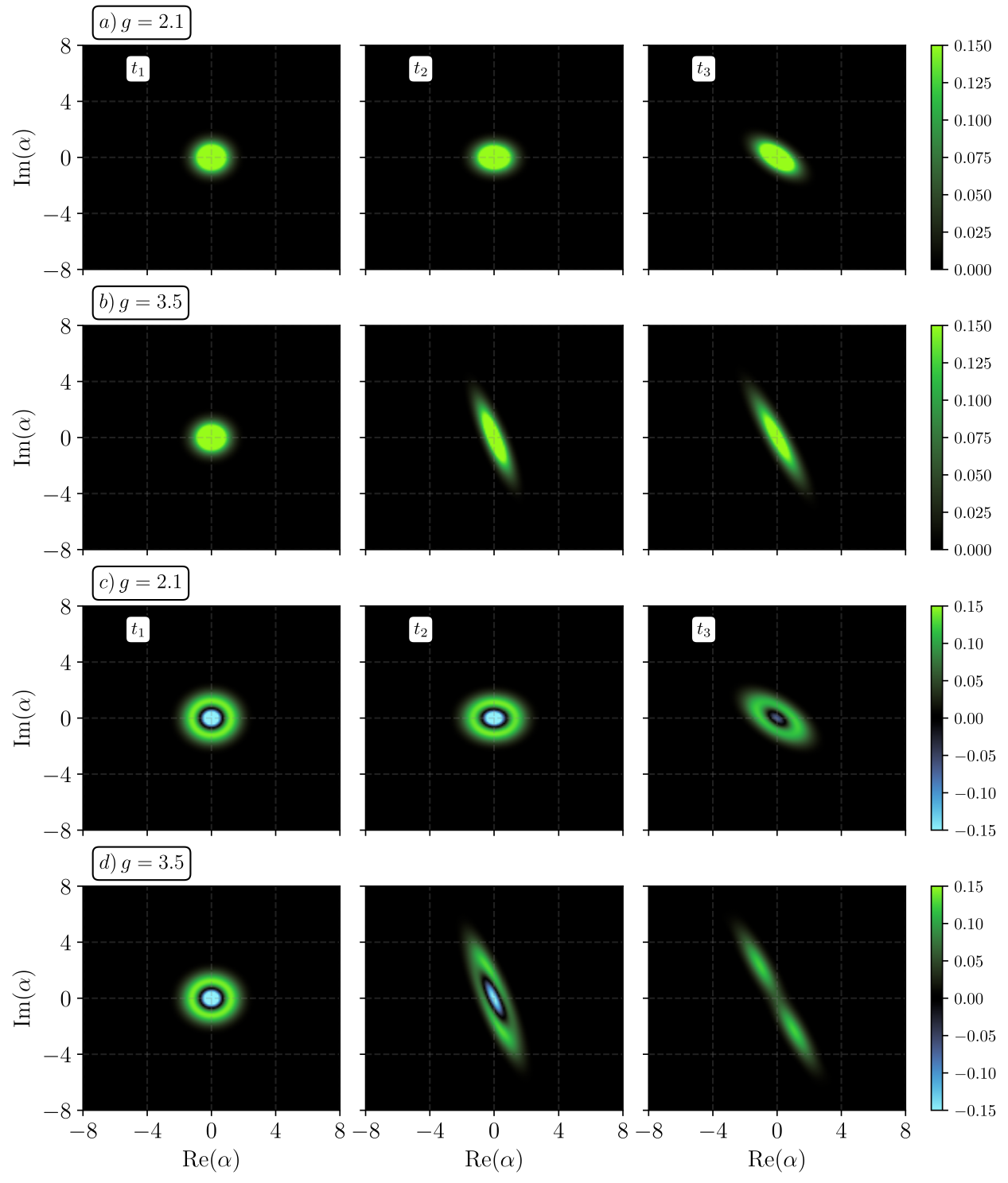}
 	\caption{Phase-space evolution of the Wigner function $W_1(\alpha)$ for the first mode, illustrating the emergence of squeezing, nonclassicality, and classical-like behavior across different dynamical regimes. Panels a) and b) correspond to the initial vacuum state $|\psi(0)\rangle = |0,0\rangle$, showing Gaussian and strictly positive Wigner distributions that undergo progressive squeezing, increasingly pronounced as the coupling $g$ approaches the transition region. Panels c) and d) correspond to the initially excited state $|\psi(0)\rangle = |1,0\rangle$, where the appearance of negative regions signals nonclassical behavior that is largely preserved during the evolution. For specific combinations of time and strong coupling (e.g., $g=3.5$), the Wigner function becomes entirely positive, suggesting an effective restoration of classicality associated with the transition to a non--harmonic-oscillator regime.}
 	\label{fig:wfunc}
\end{figure}

\pagebreak
\section{Conclusions}\label{conclu}
In this work, we revisited the effective Hamiltonian originally introduced by C.~K.~Law for a non-stationary electromagnetic cavity and analyzed its behavior under uniform mirror motion. By applying a sequence of time-independent geometric transformations, we established conditions for the unitary diagonalization of the model, thereby defining a symmetric bound on the normalized velocity. Although the resulting interval $|v/c| \lesssim 10.17$ exceeds the physically admissible regime, it provides a consistent mathematical framework that motivated a reinterpretation of the system.

We proposed an alternative photonic viewpoint in which the {\color{black}normalized velocity} $g = v/c$ is treated as an effective coupling constant between optical modes. Within this framework, the effective Hamiltonian realizes the generators of the symplectic algebra $\mathrm{Sp}(4,\mathbb{R})$, which naturally describe two-mode squeezing and linear optical transformations. This algebraic correspondence allows us to reinterpret the cavity–mirror model as a photonic platform for generating non-standard two-mode squeezed states and exploring strong-coupling regimes inaccessible in the original electrodynamic setting.

After diagonalizing the Hamiltonian, we derived analytical expressions for the mean photon number and examined statistical indicators, including the Mandel parameter, fidelity, and linear entropy. These quantities reveal transitions between sub- and super-Poissonian statistics and between pure and mixed states, indicating non-classical correlations and entanglement. The Wigner-function analysis provided complementary phase-space insight, showing how vacuum and excited initial states evolve into squeezed states whose positivity or negativity reflects the transition between classical and quantum regimes.

Our results complement the recent analysis of Ramos-Prieto et al.~\cite{RamosPrieto2023}, who explored the temporal factorization of the effective Hamiltonian in the context of uniform mirror motion. In contrast, we focus on the underlying symplectic structure and show that the same Hamiltonian can be consistently reinterpreted as a photonic analog in which the coupling parameter replaces the {\color{black}normalized velocity}. This reinterpretation broadens the model's applicability, providing a mathematically consistent and experimentally relevant framework for simulating non-stationary quantum fields using photonic systems.

The formalism presented here can be extended to include time-dependent couplings, dissipative channels, or accelerated boundary motion, paving the way for experimentally realizable analog simulations of quantum vacuum phenomena and field dynamics in engineered optical and superconducting platforms.

\section{Acknowledgments}
RA-U acknowledges financial support from the UNAM Postdoctoral Program (POSDOC) 2023-2025 and thanks ICF-UNAM for the assistance provided. J.R-A and L.M-D acknowledge partial support from the DGAPA-UNAM project IN109822. R.R.-A. thanks DGAPA-UNAM, Mexico, for support under Project No. IA101826.

\begin{appendices}

\section{Coefficients for the expectation values}\label{apA}
In this appendix, we will explicitly provide the relations for the coefficients defining the time evolution of the bosonic operators in Eqs.~\eqref{n1} and \eqref{n2}. First, let us define some auxiliary terms,
\begin{equation}
    \begin{aligned}
        &c_{j}^{(1)}(t) = \cos\left(2 F(t) \sqrt{\alpha_{j}\beta_{j}}\right),\quad c_{j}^{(2)}(t) = \sin\left(2 F(t) \sqrt{\alpha_{j}\beta_{j}}\right),\quad r_{j} = \sqrt{\frac{\alpha_{j}}{\beta_{j}}},\\
        &a_{1,1} = - \frac{\mathrm{i} g c^{2}}{8 \pi} - \frac{\mathrm{i} \pi}{2}, \qquad a_{1,2} = - \frac{\mathrm{i} \chi^{2} \xi^{2}}{2 \pi} + \frac{\mathrm{i} \chi \xi}{\pi} - \frac{\mathrm{i}}{2 \pi},\\
        &a_{1,3} = - \frac{\mathrm{i} g c^{2} \xi^{2}}{8 \pi} - \frac{\mathrm{i} \pi \xi^{2}}{2}, \qquad a_{1,4} = - \frac{\sqrt{2} \mathrm{i} \chi g c}{8 \pi} + \frac{\sqrt{2} \chi}{4},\\
        &a_{1,5} = - \frac{\sqrt{2} \mathrm{i} \chi g c \xi^{2}}{8 \pi} - \frac{\sqrt{2} \chi \xi^{2}}{2} + \frac{\sqrt{2} \mathrm{i} g c \xi}{8 \pi} + \frac{\sqrt{2} \xi}{2},\\ &a_{1,6} = - \frac{\sqrt{2} \mathrm{i} \chi^{2} \xi}{4 \pi} + \frac{\sqrt{2} \mathrm{i} \chi}{4 \pi} + \frac{\sqrt{2} \mathrm{i} g c^{2} \xi}{16 \pi} + \frac{\sqrt{2} g c \xi}{8} + \frac{\sqrt{2} \mathrm{i} \pi \xi}{2}.
    \end{aligned}
\end{equation}

\subsection{Quadrature transformation relations}\label{apAsub1}
The explicit forward and backward transformations of the quadrature operators under the composite unitary operator $\hat{U} = \hat{R}_1 \hat{R}_2 \hat{R}_3$ (Eq. 23) are given by:
\begin{equation}
\begin{aligned}
\hat{U}^\dagger \hat{x}_1 \hat{U} &= \xi \hat{p}_2 + \hat{x}_1,
\\
\hat{U} \hat{x}_1 \hat{U}^\dagger &= -\frac{g c \xi}{2} \hat{x}_2 - \xi \hat{p}_2 + (1 - \chi \xi) \hat{x}_1,
\\
\hat{U}^\dagger \hat{p}_1 \hat{U} &= -\chi \hat{x}_2 - \frac{g c \xi}{2} \hat{p}_2 - \frac{g c}{2} \hat{x}_1 + (1 - \chi \xi) \hat{p}_1,
\\
\hat{U} \hat{p}_1 \hat{U}^\dagger &= \chi \hat{x}_2 + \frac{g c}{2} \hat{x}_1 + \hat{p}_1, 
\\
\hat{U}^\dagger \hat{x}_2 \hat{U} &= \xi \hat{p}_1 + \hat{x}_2, 
\\
\hat{U} \hat{x}_2 \hat{U}^\dagger &= -\frac{g c \xi}{2} \hat{x}_1 - \xi \hat{p}_1 + (1 - \chi \xi) \hat{x}_2, 
\\ 
\hat{U}^\dagger \hat{p}_2 \hat{U} &= -\chi \hat{x}_1 - \frac{g c \xi}{2} \hat{p}_1 - \frac{g c}{2} \hat{x}_2 + (1 - \chi \xi) \hat{p}_2,
\\
\hat{U} \hat{p}_2 \hat{U}^\dagger &= \chi \hat{x}_1 + \frac{g c}{2} \hat{x}_2 + \hat{p}_2.
\end{aligned}
\end{equation}

\subsection{First mode}\label{nexp_coeffs_1}
The coefficients in the expectation value for mode 1 are given by
\begin{equation}
\begin{aligned}
\eta_{1,1} =& \chi\xi c_{2}^{(1)}(t) + (1 - \chi\xi) c_{1}^{(1)}(t) + \left(a_{1,1} r_{1}^{-1} + a_{1,2} r_{1}\right) c_{1}^{(2)}(t)
\\ & 
- \left(\frac{\mathrm{i}\chi^{2}}{2\pi}r_{2}^{-1} + a_{1,1}\xi^{2} r_{2}\right) c_{2}^{(2)}(t),\\
\eta_{1,2} =& \left(a_{1,3} r_{1}^{-1} + a_{1,2} r_{1}\right) c_{1}^{(2)}(t) 
\\ &
- \left(\frac{\mathrm{i}\chi^{2}}{2\pi}r_{2}^{-1} + a_{1,3} \xi^{2} r_{2}\right) c_{2}^{(2)}(t),\\
\eta_{1,3} = & \left(a_{1,4} r_{1}^{-1} + a_{1,5} r_{1}\right)c_{1}^{(2)}(t) - \left(a_{1,4}^{*} r_{2}^{-1} + a_{1,5}^{*} r_{2}\right) c_{2}^{(2)}(t) 
\\ &
+ a_{1,6} c_{1}^{(1)}(t) - a_{1,6}^{*} c_{2}^{(1)}(t),\\
\eta_{1,4} = & \left(a_{1,4} r_{1}^{-1} - a_{1,5}^{*} r_{1}\right) c_{1}^{(2)}(t) - \left(a_{1,4}^{*} r_{2}^{-1} - a_{1,5} r_{2}\right) c_{2}^{(2)}(t) 
\\ &
- a_{1,6}^{*} c_{1}^{(1)}(t) + a_{1,6}^{*} c_{2}^{(1)}(t).
\end{aligned}
\end{equation}

\subsection{Second mode}\label{nexp_coeffs_2}
The coefficients in the expectation value for the second mode are given by
\begin{equation}
\begin{aligned}
\eta_{2,1} =& -\left(a_{1,4}^{*} r_{1}^{-1} + a_{1,5}^{*} r_{1}\right) c_{1}^{(2)}(t) + \left(a_{1,4} r_{2}^{-1} + a_{1,5} r_{2}\right) c_{2}^{(2)}(t)
\\ &
 - a_{1,6}^{*} c_{2}^{(1)}(t) - a_{1,6} c_{1}^{(1)}(t),\\
\eta_{2,2} =& \left(a_{1,4} r_{2}^{-1} + a_{1,5} r_{2}\right) c_{1}^{(2)}(t) + \left(a_{1,4} r_{2}^{-1} + a_{1,5} r_{2}\right) c_{2}^{(2)}(t) 
\\ &
+ a_{1,6}^{*} c_{2}^{(1)}(t) - a_{1,6} c_{1}^{(1)}(t),\\
\eta_{2,3} =& \chi\xi c_{1}^{(1)}(t) + (1 - \chi\xi) c_{2}^{(1)}(t) + \left(a_{1,1} r_{2}^{-1} + a_{1,2} r_{2}\right) c_{2}^{(2)}(t) 
\\ &
- \left(\frac{\mathrm{i}\chi^{2}}{4\pi}  r_{1}^{-1} + a_{1,3} r_{1}\right) c_{1}^{(2)}(t),\\
\eta_{2,4} =& -\left[\left(a_{1,1}^{*} - \frac{v}{2}\right) r_{2}^{-1} - a_{1,2} r_{2}\right] c_{2}^{(2)}(t) - \left(\frac{\mathrm{i}\chi^{2}}{4\pi} r_{1}^{-1} - \frac{1}{8} a_{1,2}^{*} r_{1} \right) c_{1}^{(2)}(t).
    \end{aligned}
\end{equation}

{\color{black}
\section{Derivation of the single-mode Wigner functions}\label{app:wigner_derivation}
In this appendix, we provide the explicit phase-space derivation of the single-mode Wigner function $W_1(\alpha, t)$ for the first mode under the symplectic dynamics generated by the effective Hamiltonian.

The reduced Wigner function $W_1(\alpha, t)$ for mode 1 is defined as the two-dimensional Fourier transform of the single-mode quantum characteristic function $\chi_1(\beta, t)$:
\begin{equation}
W_1(\alpha, t) = \frac{1}{\pi^2} \int d^2\beta \, \chi_1(\beta, t) \, e^{\alpha \beta^* - \alpha^* \beta},
\label{eq:Wigner_def}
\end{equation}
where $\chi_1(\beta, t)$ is given by the expectation value of the single-mode Glauber displacement operator $\hat{D}_1(\beta) = \exp(\beta \hat{a}_1^\dagger - \beta^* \hat{a}_1)$:
\begin{equation}
\chi_1(\beta, t) = \text{Tr} \left\{ \hat{\rho}_{1,2}(t) \left( \hat{D}_1(\beta) \otimes \hat{I}_2 \right) \right\} = \langle \psi(0) | \hat{U}^\dagger(t) \hat{D}_1(\beta) \hat{U}(t) | \psi(0) \rangle.
\label{eq:char_def}
\end{equation}
Here, $\hat{U}(t) = \hat{R}_1 \hat{R}_2 \hat{R}_3 e^{-i F(t) \hat{H}_R} \hat{R}_3^\dagger \hat{R}_2^\dagger \hat{R}_1^\dagger$ represents the composite unitary time-evolution operator defined in Eq.~\eqref{eq0110}, with $F(t)$ the properly time-integrated exponent introduced in Eq.~(21).

In the Heisenberg picture, the transformed displacement operator takes the form $\hat{U}^\dagger(t) \hat{D}_1(\beta) \hat{U}(t) = \exp \left[ \beta \hat{a}_{1}^\dagger(t) - \beta^* \hat{a}_1(t) \right]$. According to the linear Bogoliubov map established in Eqs.~\eqref{eq0250}--\eqref{eq0270}, the time-evolved annihilation operator $\hat{a}_1(t)$ expands in terms of the initial mode operators $\{\hat{a}_1, \hat{a}_1^\dagger, \hat{a}_2, \hat{a}_2^\dagger\}$ as
\begin{equation}
\hat{a}_1(t) = \eta_{1,1} \hat{a}_1 + \eta_{1,2} \hat{a}_1^\dagger + \eta_{1,3} \hat{a}_2 + \eta_{1,4} \hat{a}_2^\dagger,
\label{eq:a1_t_bogo}
\end{equation}
where $\hat{a}_k \equiv \hat{a}_k(0)$. Substituting Eq.~\eqref{eq:a1_t_bogo} into the exponent of the displacement operator yields
\begin{equation}
\beta \hat{a}_1^\dagger(t) - \beta^* \hat{a}_1(t) = \left( \xi_1 \hat{a}_1^\dagger - \xi_1^* \hat{a}_1 \right) + \left( \xi_2 \hat{a}_2^\dagger - \xi_2^* \hat{a}_2 \right),
\label{eq:exp_split}
\end{equation}
with the effective displacement amplitudes
\begin{equation}
\xi_1(\beta, t) = \beta \eta_{1,1}^* - \beta^* \eta_{1,2}, \qquad \xi_2(\beta, t) = \beta \eta_{1,3}^* - \beta^* \eta_{1,4}.
\label{eq:xi_defs}
\end{equation}
Since operators belonging to distinct modes commute, the transformed displacement operator factorizes cleanly into a product of single-mode displacement operators,
\begin{equation}
\hat{U}^\dagger(t) \hat{D}_1(\beta) \hat{U}(t) = \hat{D}_1(\xi_1) \otimes \hat{D}_2(\xi_2).
\label{eq:disp_factor}
\end{equation}

\subsection{General Fock-state characteristic function}\label{app:general_fock}
Rather than treating each initial Fock state separately, it is more economical to use the well-known displaced number-state overlap \cite{CahillGlauber1969, CahillGlauber1969b}
\begin{equation}
\langle n | \hat{D}(\xi) | n \rangle = L_n\!\left(|\xi|^2\right) e^{-\frac{1}{2}|\xi|^2},
\label{eq:displaced_number_overlap}
\end{equation}
with $L_n$ the $n$-th Laguerre polynomial. For an arbitrary initial product Fock state $|\psi(0)\rangle = |n,m\rangle$, Eq.~\eqref{eq:disp_factor} then gives, in one stroke,
\begin{equation}
\chi_1^{(n,m)}(\beta, t) = L_n\!\left(|\xi_1|^2\right) L_m\!\left(|\xi_2|^2\right) \exp\left[-\tfrac{1}{2}\left(|\xi_1|^2+|\xi_2|^2\right)\right].
\label{eq:chi_general}
\end{equation}
Both cases considered below follow immediately from Eq.~\eqref{eq:chi_general}, since $L_0(x)=1$ and $L_1(x)=1-x$: the vacuum sector is the $n=m=0$ corollary, and the single-photon sector the $n=1,m=0$ corollary. No separate re-derivation is required.

We record the quadratic expansion of $|\xi_1|^2$ once, as it is used by both sectors below. Expanding Eq.~\eqref{eq:xi_defs} directly,
\begin{equation}
|\xi_1(\beta,t)|^2 = \left(|\eta_{1,1}|^2 + |\eta_{1,2}|^2\right)|\beta|^2 \;-\; \eta_{1,1}^*\eta_{1,2}^*\,\beta^2 \;-\; \eta_{1,1}\eta_{1,2}\,\beta^{*2},
\label{eq:xi1_sq_expanded}
\end{equation}
and analogously for $|\xi_2|^2$ with $\{\eta_{1,1},\eta_{1,2}\}\to\{\eta_{1,3},\eta_{1,4}\}$.

\subsection{Vacuum initial state: $n=m=0$}
Setting $n=m=0$ in Eq.~\eqref{eq:chi_general} gives the Gaussian characteristic function
\begin{equation}
\chi_1^{(0,0)}(\beta, t) = \exp \left( -\frac{1}{2} |\xi_1|^2 - \frac{1}{2} |\xi_2|^2 \right) = \exp\!\left[-A(t)|\beta|^2 - \tfrac{1}{2}B(t)\beta^2 - \tfrac{1}{2}B^*(t)\beta^{*2}\right],
\label{eq:chi_vac}
\end{equation}
where, summing the contributions of Eq.~\eqref{eq:xi1_sq_expanded} and its mode-2 analogue,
\begin{equation}
A(t) = \frac{1}{2} \sum_{m=1}^4 |\eta_{1,m}|^2 = \langle \hat{n}_1(t) \rangle_{0,0} + \frac{1}{2}, \qquad B(t) = - \left( \eta_{1,1}^* \eta_{1,2}^* + \eta_{1,3}^* \eta_{1,4}^* \right).
\end{equation}
Evaluating the 2D Gaussian Fourier integral in Eq.~\eqref{eq:Wigner_def} leads directly to a squeezed Gaussian distribution in phase space,
\begin{equation}
W_1^{(0,0)}(\alpha, t) = \frac{1}{\pi \sqrt{\det \boldsymbol{\sigma}_1(t)}} \exp \left( -\frac{1}{2} \boldsymbol{\bar{r}}^T \boldsymbol{\sigma}_1^{-1}(t) \boldsymbol{\bar{r}} \right),
\label{eq:Wigner_vac_res}
\end{equation}
where $\boldsymbol{\bar{r}} = (\text{Re}(\alpha), \text{Im}(\alpha))^T$ and $\boldsymbol{\sigma}_1(t)$ is the $2\times2$ covariance matrix of the quadratures of mode 1,
\begin{equation}
\boldsymbol{\sigma}_1(t) = \begin{pmatrix} \langle \Delta \hat{x}_1^2 \rangle & \frac{1}{2} \langle \{ \Delta \hat{x}_1, \Delta \hat{p}_1 \} \rangle \\ \frac{1}{2} \langle \{ \Delta \hat{x}_1, \Delta \hat{p}_1 \} \rangle & \langle \Delta \hat{p}_1^2 \rangle \end{pmatrix}.
\end{equation}
Equation~\eqref{eq:Wigner_vac_res} shows that an initial vacuum state evolves into a strictly positive, Gaussian squeezed thermal state.

\subsection{Single-photon initial state: $n=1, m=0$}
Setting $n=1,m=0$ in Eq.~\eqref{eq:chi_general} gives, using $L_1(x)=1-x$,
\begin{equation}
\chi_1^{(1,0)}(\beta, t) = \left(1-|\xi_1(\beta,t)|^2\right)\chi_1^{(0,0)}(\beta,t).
\label{eq:chi_10}
\end{equation}
To Fourier-transform the extra polynomial factor, we first fix, unambiguously, the differential representation of $\beta,\beta^*$ under the transform in Eq.~\eqref{eq:Wigner_def}. Differentiating the kernel gives $\partial_\alpha e^{\alpha\beta^*-\alpha^*\beta} = \beta^*e^{\alpha\beta^*-\alpha^*\beta}$ and $\partial_{\alpha^*}e^{\alpha\beta^*-\alpha^*\beta} = -\beta\,e^{\alpha\beta^*-\alpha^*\beta}$; since $\chi_1$ decays at infinity, integrating Eq.~\eqref{eq:Wigner_def} by parts term by term gives the operator correspondence
\begin{equation}
\beta\,\chi_1(\beta,t) \;\longleftrightarrow\; -\frac{\partial}{\partial\alpha^*}\,W_1(\alpha,t), \qquad
\beta^*\,\chi_1(\beta,t) \;\longleftrightarrow\; \frac{\partial}{\partial\alpha}\,W_1(\alpha,t).
\label{eq:bopp_shift}
\end{equation}
Applying Eq.~\eqref{eq:bopp_shift} term by term to the expansion of $|\xi_1|^2$ already recorded in Eq.~\eqref{eq:xi1_sq_expanded} — i.e. $|\beta|^2=\beta\beta^*\to-\partial_\alpha\partial_{\alpha^*}$, $\beta^2\to\partial_{\alpha^*}^2$, $\beta^{*2}\to\partial_\alpha^2$ — gives the corresponding phase-space differential operator
\begin{equation}
\mathcal{P}\!\left(\frac{\partial}{\partial \alpha}, \frac{\partial}{\partial \alpha^*}\right) = \left(|\eta_{1,1}|^2 + |\eta_{1,2}|^2\right) \left(-\frac{\partial^2}{\partial \alpha \, \partial \alpha^*}\right) - \eta_{1,1} \eta_{1,2} \frac{\partial^2}{\partial \alpha^2} - \eta_{1,1}^* \eta_{1,2}^* \frac{\partial^2}{\partial \alpha^{*2}},
\label{eq:P_operator}
\end{equation}
so that Eq.~\eqref{eq:chi_10} transforms into
\begin{equation}
W_1^{(1,0)}(\alpha, t) = \left[ 1 - \mathcal{P}\left(\frac{\partial}{\partial \alpha}, \frac{\partial}{\partial \alpha^*}\right) \right] W_1^{(0,0)}(\alpha, t).
\label{eq:W10_operator}
\end{equation}
Executing these derivatives on the Gaussian of Eq.~\eqref{eq:Wigner_vac_res} yields the non-Gaussian expression
\begin{equation}
W_{1}^{(1,0)}(\alpha, t) = W_1^{(0,0)}(\alpha, t) \cdot \mathcal{Q}(\alpha, \alpha^*, t),
\label{eq:Wigner_excited_res}
\end{equation}
where $\mathcal{Q}(\alpha, \alpha^*, t)$ is a second-order polynomial in the phase-space coordinates $\text{Re}(\alpha)$ and $\text{Im}(\alpha)$, with the squeezing-sensitive terms now correctly weighted by the unconjugated pair $\eta_{1,1}\eta_{1,2}$ and its conjugate. The polynomial $\mathcal{Q}(\alpha, \alpha^*, t)$ introduces a negative region near the origin, revealing the nonclassical character of the single-photon excitation. As time evolves, the dynamic shearing of the covariance matrix $\boldsymbol{\sigma}_1(t)$ governs the reshaping and transient loss/restoration of negativity illustrated in Fig.~6.
}
\end{appendices}

\bibliographystyle{unsrt}
\bibliography{references}

\end{document}